\documentstyle[10pt,epsf,epsfig,hangcaption,xspace,amssymb,amsfonts,amsmath,amsthm,cite,
dp_delphititle,lineno]{dp_delphi}
\makeindex
\def\DpPaperGroup{EP}
\def\DpPaperRef{2003-070}
\def\DpDate{21 July 2003}
\def\DpAuthors{DELPHI Collaboration}
\def\DpSubmit{(Eur. Phys. J. C36 (2004) 283-296)}
\def\DpTitle{{A Precise Measurement of the Tau Lifetime}}
\def\DpComment{ }
\def\DpEMail{ }

\newcommand{\msub}[1]{{\mbox{\scriptsize #1}}}
\def\mtau{m_\tau}
\def\mtau2{m^2_{\tau}}

\def\elec{\rm e}
\def\ee{ {\rm e}^+ {\rm e}^-}
\def\tt{ \tau^+\tau^-}
\def\mm{ \mu^+\mu^-}

\def\qq{ {\rm q} \bar{\rm q}}

\def\eemm{\ee \rightarrow \mm}
\def\bhab{\ee \rightarrow \ee}
\def\eeha{\ee \rightarrow \qq}

\newcommand{\aone}{{\rm a}_1}
\def\a1nu{\aone\nu}

\def\s2thw{\sin^{2}\theta^{\mbox{\scriptsize lept}}_{\mbox{\scriptsize
eff}}}

\newcommand{\nut}{\nu_{\tau}}
\newcommand{\num}{\nu_{\mu}}
\newcommand{\nue}{\nu_{\elec}}
\newcommand{\Tauto}{\tau^- \rightarrow}

\newcommand{\MU}{\mu^- \bar{\num} \nut}
\newcommand{\TMU}{\Tauto \MU}
\newcommand{\EL}{\elec^- \bar{\nue} \nut}
\newcommand{\TEL}{\Tauto \EL}

\begin{document}
\makeatletter
\newcount\@tempcntc
\def\@citex[#1]#2{\if@filesw\immediate\write\@auxout{\string\citation{#2}}\fi
  \@tempcnta\z@\@tempcntb\m@ne\def\@citea{}\@cite{\@for\@citeb:=#2\do
    {\@ifundefined
       {b@\@citeb}{\@citeo\@tempcntb\m@ne\@citea\def\@citea{,}{\bf ?}\@warning
       {Citation `\@citeb' on page \thepage \space undefined}}%
    {\setbox\z@\hbox{\global\@tempcntc0\csname b@\@citeb\endcsname\relax}%
     \ifnum\@tempcntc=\z@ \@citeo\@tempcntb\m@ne
       \@citea\def\@citea{,}\hbox{\csname b@\@citeb\endcsname}%
     \else
      \advance\@tempcntb\@ne
      \ifnum\@tempcntb=\@tempcntc
      \else\advance\@tempcntb\m@ne\@citeo
      \@tempcnta\@tempcntc\@tempcntb\@tempcntc\fi\fi}}\@citeo}{#1}}
\def\@citeo{\ifnum\@tempcnta>\@tempcntb\else\@citea\def\@citea{,}%
  \ifnum\@tempcnta=\@tempcntb\the\@tempcnta\else
   {\advance\@tempcnta\@ne\ifnum\@tempcnta=\@tempcntb \else \def\@citea{--}\fi
    \advance\@tempcnta\m@ne\the\@tempcnta\@citea\the\@tempcntb}\fi\fi}
 
\makeatother
\begin{titlepage}
\pagenumbering{roman}
\CERNpreprint{\DpPaperGroup}{\DpPaperRef} 
\date{{\small\DpDate}} 
\title{\DpTitle} 
\address{\DpAuthors} 
\begin{shortabs} 
\noindent

\def\ee{ \mbox{\rm e}^+ \mbox{\rm e}^-}
\def\tt{ \tau^+\tau^-}
The tau lepton lifetime has been measured with the ${\ee \rightarrow \tt}$
events collected by the DELPHI detector at LEP in the years 1991-1995.
Three different
methods have been exploited, using both one-prong and three-prong $\tau$
decay channels.
Two measurements have been made using events in which both taus decay to
a single charged particle.  Combining these measurements gave
$\tau_\tau \mbox{(1 prong)} = 291.8\pm 2.3_{stat} \pm 1.5_{sys}  \mbox{ fs}$.
A third measurement using taus which decayed to three charged particles
yielded
$\tau_\tau \mbox{(3 prong)} = 288.6\pm 2.4_{stat} \pm
1.3_{sys} \mbox{ fs}. $
These were combined with previous DELPHI results
to measure the tau lifetime, using the full LEP1 data sample, to be
$\tau_\tau = 290.9\pm 1.4_{stat} \pm 1.0_{sys} \mbox{\ fs}$.

\end{shortabs}
\vfill
\begin{center}
\DpSubmit \ \\ 
\DpComment \ \\
\DpEMail \ \\
\end{center}
\vfill
\clearpage
\headsep 10.0pt
\addtolength{\textheight}{10mm}
\addtolength{\footskip}{-5mm}
\begingroup
%
\newcommand{\DpName}[2]{\hbox{#1$^{\ref{#2}}$},\hfill}
\newcommand{\DpNameTwo}[3]{\hbox{#1$^{\ref{#2},\ref{#3}}$},\hfill}
\newcommand{\DpNameThree}[4]{\hbox{#1$^{\ref{#2},\ref{#3},\ref{#4}}$},\hfill}
\newskip\Bigfill \Bigfill = 0pt plus 1000fill
\newcommand{\DpNameLast}[2]{\hbox{#1$^{\ref{#2}}$}\hspace{\Bigfill}}
%
\footnotesize
\noindent
\DpName{J.Abdallah}{LPNHE}
\DpName{P.Abreu}{LIP}
\DpName{W.Adam}{VIENNA}
\DpName{P.Adzic}{DEMOKRITOS}
\DpName{T.Albrecht}{KARLSRUHE}
\DpName{T.Alderweireld}{AIM}
\DpName{R.Alemany-Fernandez}{CERN}
\DpName{T.Allmendinger}{KARLSRUHE}
\DpName{P.P.Allport}{LIVERPOOL}
\DpName{U.Amaldi}{MILANO2}
% ...
\normalsize
\endgroup
\titlefoot{Department of Physics and Astronomy, Iowa State
     University, Ames IA 50011-3160, USA
    \label{AMES}}
\titlefoot{Physics Department, Universiteit Antwerpen,
     Universiteitsplein 1, B-2610 Antwerpen, Belgium \\
     \indent~~and IIHE, ULB-VUB,
     Pleinlaan 2, B-1050 Brussels, Belgium \\
     \indent~~and Facult\'e des Sciences,
     Univ. de l'Etat Mons, Av. Maistriau 19, B-7000 Mons, Belgium
    \label{AIM}}
\titlefoot{Physics Laboratory, University of Athens, Solonos Str.
     104, GR-10680 Athens, Greece
    \label{ATHENS}}
\titlefoot{Department of Physics, University of Bergen,
     All\'egaten 55, NO-5007 Bergen, Norway
    \label{BERGEN}}
\titlefoot{Dipartimento di Fisica, Universit\`a di Bologna and INFN,
     Via Irnerio 46, IT-40126 Bologna, Italy
    \label{BOLOGNA}}
\titlefoot{Centro Brasileiro de Pesquisas F\'{\i}sicas, rua Xavier Sigaud 150,
     BR-22290 Rio de Janeiro, Brazil \\
     \indent~~and Depto. de F\'{\i}sica, Pont. Univ. Cat\'olica,
     C.P. 38071 BR-22453 Rio de Janeiro, Brazil \\
     \indent~~and Inst. de F\'{\i}sica, Univ. Estadual do Rio de Janeiro,
     rua S\~{a}o Francisco Xavier 524, Rio de Janeiro, Brazil
    \label{BRASIL}}
\titlefoot{Coll\`ege de France, Lab. de Physique Corpusculaire, IN2P3-CNRS,
     FR-75231 Paris Cedex 05, France
    \label{CDF}}
\titlefoot{CERN, CH-1211 Geneva 23, Switzerland
    \label{CERN}}
\titlefoot{Institut de Recherches Subatomiques, IN2P3 - CNRS/ULP - BP20,
     FR-67037 Strasbourg Cedex, France
    \label{CRN}}
\titlefoot{Now at DESY-Zeuthen, Platanenallee 6, D-15735 Zeuthen, Germany
    \label{DESY}}
\titlefoot{Institute of Nuclear Physics, N.C.S.R. Demokritos,
     P.O. Box 60228, GR-15310 Athens, Greece
    \label{DEMOKRITOS}}
\titlefoot{FZU, Inst. of Phys. of the C.A.S. High Energy Physics Division,
     Na Slovance 2, CZ-180 40, Praha 8, Czech Republic
    \label{FZU}}
\titlefoot{Dipartimento di Fisica, Universit\`a di Genova and INFN,
     Via Dodecaneso 33, IT-16146 Genova, Italy
    \label{GENOVA}}
\titlefoot{Institut des Sciences Nucl\'eaires, IN2P3-CNRS, Universit\'e
     de Grenoble 1, FR-38026 Grenoble Cedex, France
    \label{GRENOBLE}}
\titlefoot{Helsinki Institute of Physics, P.O. Box 64,
     FIN-00014 University of Helsinki, Finland
    \label{HELSINKI}}
\titlefoot{Joint Institute for Nuclear Research, Dubna, Head Post
     Office, P.O. Box 79, RU-101 000 Moscow, Russian Federation
    \label{JINR}}
\titlefoot{Institut f\"ur Experimentelle Kernphysik,
     Universit\"at Karlsruhe, Postfach 6980, DE-76128 Karlsruhe,
     Germany
    \label{KARLSRUHE}}
\titlefoot{Institute of Nuclear Physics PAN,Ul. Radzikowskiego 152,
     PL-31142 Krakow, Poland
    \label{KRAKOW1}}
\titlefoot{Faculty of Physics and Nuclear Techniques, University of Mining
     and Metallurgy, PL-30055 Krakow, Poland
    \label{KRAKOW2}}
\titlefoot{Universit\'e de Paris-Sud, Lab. de l'Acc\'el\'erateur
     Lin\'eaire, IN2P3-CNRS, B\^{a}t. 200, FR-91405 Orsay Cedex, France
    \label{LAL}}
\titlefoot{School of Physics and Chemistry, University of Lancaster,
     Lancaster LA1 4YB, UK
    \label{LANCASTER}}
\titlefoot{LIP, IST, FCUL - Av. Elias Garcia, 14-$1^{o}$,
     PT-1000 Lisboa Codex, Portugal
    \label{LIP}}
\titlefoot{Department of Physics, University of Liverpool, P.O.
     Box 147, Liverpool L69 3BX, UK
    \label{LIVERPOOL}}
\titlefoot{Dept. of Physics and Astronomy, Kelvin Building,
     University of Glasgow, Glasgow G12 8QQ
    \label{GLASGOW}}
\titlefoot{LPNHE, IN2P3-CNRS, Univ.~Paris VI et VII, Tour 33 (RdC),
     4 place Jussieu, FR-75252 Paris Cedex 05, France
    \label{LPNHE}}
\titlefoot{Department of Physics, University of Lund,
     S\"olvegatan 14, SE-223 63 Lund, Sweden
    \label{LUND}}
\titlefoot{Universit\'e Claude Bernard de Lyon, IPNL, IN2P3-CNRS,
     FR-69622 Villeurbanne Cedex, France
    \label{LYON}}
\titlefoot{Dipartimento di Fisica, Universit\`a di Milano and INFN-MILANO,
     Via Celoria 16, IT-20133 Milan, Italy
    \label{MILANO}}
\titlefoot{Dipartimento di Fisica, Univ. di Milano-Bicocca and
     INFN-MILANO, Piazza della Scienza 2, IT-20126 Milan, Italy
    \label{MILANO2}}
\titlefoot{IPNP of MFF, Charles Univ., Areal MFF,
     V Holesovickach 2, CZ-180 00, Praha 8, Czech Republic
    \label{NC}}
\titlefoot{NIKHEF, Postbus 41882, NL-1009 DB
     Amsterdam, The Netherlands
    \label{NIKHEF}}
\titlefoot{National Technical University, Physics Department,
     Zografou Campus, GR-15773 Athens, Greece
    \label{NTU-ATHENS}}
\titlefoot{Physics Department, University of Oslo, Blindern,
     NO-0316 Oslo, Norway
    \label{OSLO}}
\titlefoot{Dpto. Fisica, Univ. Oviedo, Avda. Calvo Sotelo
     s/n, ES-33007 Oviedo, Spain
    \label{OVIEDO}}
\titlefoot{Department of Physics, University of Oxford,
     Keble Road, Oxford OX1 3RH, UK
    \label{OXFORD}}
\titlefoot{Dipartimento di Fisica, Universit\`a di Padova and
     INFN, Via Marzolo 8, IT-35131 Padua, Italy
    \label{PADOVA}}
\titlefoot{Rutherford Appleton Laboratory, Chilton, Didcot
     OX11 OQX, UK
    \label{RAL}}
\titlefoot{Dipartimento di Fisica, Universit\`a di Roma II and
     INFN, Tor Vergata, IT-00173 Rome, Italy
    \label{ROMA2}}
\titlefoot{Dipartimento di Fisica, Universit\`a di Roma III and
     INFN, Via della Vasca Navale 84, IT-00146 Rome, Italy
    \label{ROMA3}}
\titlefoot{DAPNIA/Service de Physique des Particules,
     CEA-Saclay, FR-91191 Gif-sur-Yvette Cedex, France
    \label{SACLAY}}
\titlefoot{Instituto de Fisica de Cantabria (CSIC-UC), Avda.
     los Castros s/n, ES-39006 Santander, Spain
    \label{SANTANDER}}
\titlefoot{Inst. for High Energy Physics, Serpukov
     P.O. Box 35, Protvino, (Moscow Region), Russian Federation
    \label{SERPUKHOV}}
\titlefoot{J. Stefan Institute, Jamova 39, SI-1000 Ljubljana, Slovenia
     and Laboratory for Astroparticle Physics,\\
     \indent~~Nova Gorica Polytechnic, Kostanjeviska 16a, SI-5000 Nova Gorica, Slovenia, \\
     \indent~~and Department of Physics, University of Ljubljana,
     SI-1000 Ljubljana, Slovenia
    \label{SLOVENIJA}}
\titlefoot{Fysikum, Stockholm University,
     Box 6730, SE-113 85 Stockholm, Sweden
    \label{STOCKHOLM}}
\titlefoot{Dipartimento di Fisica Sperimentale, Universit\`a di
     Torino and INFN, Via P. Giuria 1, IT-10125 Turin, Italy
    \label{TORINO}}
\titlefoot{INFN,Sezione di Torino, and Dipartimento di Fisica Teorica,
     Universit\`a di Torino, Via P. Giuria 1,\\
     \indent~~IT-10125 Turin, Italy
    \label{TORINOTH}}
\titlefoot{Dipartimento di Fisica, Universit\`a di Trieste and
     INFN, Via A. Valerio 2, IT-34127 Trieste, Italy \\
     \indent~~and Istituto di Fisica, Universit\`a di Udine,
     IT-33100 Udine, Italy
    \label{TU}}
\titlefoot{Univ. Federal do Rio de Janeiro, C.P. 68528
     Cidade Univ., Ilha do Fund\~ao
     BR-21945-970 Rio de Janeiro, Brazil
    \label{UFRJ}}
\titlefoot{Department of Radiation Sciences, University of
     Uppsala, P.O. Box 535, SE-751 21 Uppsala, Sweden
    \label{UPPSALA}}
\titlefoot{IFIC, Valencia-CSIC, and D.F.A.M.N., U. de Valencia,
     Avda. Dr. Moliner 50, ES-46100 Burjassot (Valencia), Spain
    \label{VALENCIA}}
\titlefoot{Institut f\"ur Hochenergiephysik, \"Osterr. Akad.
     d. Wissensch., Nikolsdorfergasse 18, AT-1050 Vienna, Austria
    \label{VIENNA}}
\titlefoot{Inst. Nuclear Studies and University of Warsaw, Ul.
     Hoza 69, PL-00681 Warsaw, Poland
    \label{WARSZAWA}}
\titlefoot{Fachbereich Physik, University of Wuppertal, Postfach
     100 127, DE-42097 Wuppertal, Germany
    \label{WUPPERTAL}}
\addtolength{\textheight}{-10mm}
\addtolength{\footskip}{5mm}
\clearpage
\headsep 30.0pt
\end{titlepage}
%
\pagenumbering{arabic} 
\setcounter{footnote}{0} %
\large
%
 
\section{Introduction}
\label{s:intro}
 
The tau lepton is a fundamental constituent of the Standard Model
and its lifetime can be used to test
the model's predictions. In particular, lepton
universality can be probed using the relationships
\begin{eqnarray}
      \tau_{\tau} & = & \tau_{\mu}
      \left( \frac{g_{\mu}}{g_{\tau}} \right)^{2}
      \left( \frac{{\rm m}_{\mu}}{{\rm m}_{\tau}}
 \right)^{5} \cdot \mbox{BR} \left( \tau^- \to e^-\bar{\nu_{e}}
      \nu_{\tau}\right)
      \cdot \frac{f(m^2_e/m^2_\mu)r^\mu_{RC}}{f(m^2_e/m^2_\tau)r^\tau_{RC}}, 
\label{eqn:life} \\
      \tau_{\tau} & = &\tau_{\mu}
      \left( \frac{g_{e}}{g_{\tau}} \right)^{2}
      \left( \frac{{\rm m}_{\mu}}{{\rm m}_{\tau}}
 \right)^{5} \cdot \mbox{BR} \left( \tau^- \to \mu^-\bar{\nu_{\mu}}
      \nu_{\tau}\right)
      \cdot \frac{f(m^2_e/m^2_\mu)r^\mu_{RC}}{f(m^2_\mu/m^2_\tau)r^\tau_{RC}}, 
\label{eqn:life1}
\end{eqnarray}
where $\tau_{\mu,{\tau}}$ and m$_{\mu,{\tau}}$ are the lifetimes and masses
of the muon and tau lepton, $g_{e,\mu,{\tau}}$ are the
coupling constants to the $W^{\pm}$
for the electron, muon and tau respectively, $f$ are phase space factors and 
$r^{\mu,\tau}_{RC}$ are radiative corrections to the decay 
widths~\cite{theory}. 
To the precision with which the tau lifetime and branching ratios can be
measured, $f(m^2_\mu/m^2_\tau)=0.9726$ while 
$f(m^2_e/m^2_\mu)$ and $f(m^2_e/m^2_\tau)$ are 1.000;
the electroweak radiative corrections $r^\mu_{RC}$ and $r^\tau_{RC}$ 
amount to 0.9956 and 0.9960 respectively.

The lifetime measurements presented here were performed with
the data taken by the DELPHI experiment at the LEP electron-positron collider 
at centre-of-mass energies of the $\ee$ system around 91~GeV, where
tau leptons were pair-produced through the decay of the Z boson.
As in previous measurements~\cite{DELPHI95}, the three-layer 
silicon microvertex detector~\cite{micro} and its excellent
spatial resolution
were the key to achieving the precision on track measurements necessary to 
determine the short tau decay distance.
 
Three techniques were used to measure the lifetime depending on the
final-state topology of the event.
In the channel in which a tau decayed into a final state containing
three charged particles (3-prong decays), 
it was possible to reconstruct the decay
vertex and measure the decay distance from the centre of the
interaction region of the LEP beams. An analysis of the complete sample
of such decays collected by DELPHI from 1991 to 1995 was performed.
When a tau decayed into final states with only 
one charged particle (one-prong decays),
the lifetime information was contained in the impact parameter of that
particle with respect to the centre of the interaction region.
Two complementary methods, similar to the ones used on 1991--1993
data~\cite{DELPHI95,DELPHI91} were applied to the data collected during
the 1994 and 1995 LEP running.  These methods exploited the
correlation between the impact parameters of the two charged particles in
two one-prong tau decays.

All these methods measured the tau decay length.
The conversion to a lifetime used the Lorentz boost parameter $\gamma\beta$
which was estimated from the tau mass 
$m_\tau=1776.99^{+0.29}_{-0.26}$~MeV/c$^2$~\cite{PDG02} and the energy of the LEP beams. 

The Monte Carlo program KORALZ~4.0 \cite{jadach}, together with
the TAUOLA~2.5 \cite{tauola} library were used to model tau-pair production
and decay. 
Backgrounds were studied using several generators:
DYMU3 \cite{dymu3}   for  $\eemm$  events;   BABAMC \cite{babamc}  and
BHWIDE \cite{bhwide}  for $\bhab$ events;  JETSET~7.3 with specially tuned
fragmentation parameters~\cite{jetset}
for $\eeha$ events;  BDK~\cite{berends} for reactions with
four leptons in the final state, including two-photon events where
one or two  e$^+$ or e$^-$ were not observed  in the detector; and 
TWOGAM~\cite{twogam} for $\ee\to(\ee)\qq$ events.
The generated events were interfaced to
a detailed model of the detector response~\cite{delsim}
and  reconstructed  with the  same
program  as the  real data.
Separate samples were produced corresponding to the detector
configurations in different years.
 
The DELPHI detector is described in \cite{detect}.
This analysis used the charged particle tracking system
covering the polar angle range $|\cos\theta|<0.73$.
This consisted  of four detectors in a 1.2 T solenoidal
magnetic field.
 
\begin{itemize}
 
\item
The Microvertex Detector (VD) was a three-layer silicon
vertex detector,
which provided an $R\phi$\footnote{$R$, $\phi$ and
$z$ define a cylindrical coordinate system, $+z$ being coincident with the
electron beam and $R$ and $\phi$ in the plane transverse to the beam.
The angle $\theta$ is the polar angle defined
with respect to the $z$ axis.}
precision of 7.6~$\mu$m and a two-track separation of 100~$\mu$m. 
In 1994 an upgraded version with two of the three layers equipped
 with double-sided detectors was installed, providing a $z$ precision
of 9~$\mu$m for tracks perpendicular to the beam direction.

\item
The Inner Detector (ID) was a gas detector with a jet-chamber geometry.
It measured up to 24 $R\phi$ coordinates per track, 
yielding a track element with an
$R\phi$ precision of 50~$\mu$m.
\item
The Time Projection Chamber (TPC) was the main tracking detector
of DELPHI, situated between radii of 30 cm and 120 cm.
Up to 16 points per track produced a track element with
an $R\phi$ precision of 250~$\mu$m.
\item
The Outer Detector (OD) consisted of 24 modules containing 5 layers
of drift tubes operating in limited streamer mode and situated at
a radius of 2~m. Charged particles produced track elements with
300~$\mu$m precision in $R\phi$.
\end{itemize}
 
The most important figure of merit in the tau lifetime measurement is
the precision on the impact parameter defined as
the distance of closest 
approach of a track extrapolated to the centre of the interaction region.
The impact parameter was given the same sign as the $z$ component 
of $\vec{d} \times \vec{p}_T$
where $\vec{d}$ is the projection on the $R\phi$ plane
of the vector from the centre of the interaction region to the point of
closest approach and $\vec{p}_T$ the projection on the same plane of the
particle momentum.

The precision on the impact parameter was extensively analyzed in 
the detailed study
of tracking uncertainties for the DELPHI 
measurement of the Z decay width into pairs of 
$b\bar{b}$ quarks~\cite{gammabb}.
Typically these errors correspond to an uncertainty on the impact 
parameter of 
\begin{equation}
  \sigma_\msub{track} = 20 \mbox{\ $\mu$m} \oplus 
                    \frac{65 \mbox{\ $\mu$m}}{p\mbox{[GeV/c]}\sin^{3/2}\theta}
\end{equation}
where $\oplus$ indicates the sum in quadrature.

Nevertheless this estimation was not fully satisfactory to describe the
impact parameter resolution in tau decays for two 
reasons.
Firstly, 
the topology of b events differed from that of tau decays. 
Ambiguities in the reconstruction of 
overlapping tracks degraded the resolution in b events.
This did not affect one-prong tau decays but was an even more
severe problem in three-prong decays because of the 
small opening angle of the tau decay products.
Secondly, 
particles from tau decays contain a considerably higher fraction of
muons and electrons than particles from b-decays. 
This affected the tracking precision in different ways since 
muons had no hadronic scattering in the detector material 
leading to a better tracking precision, 
while electrons were strongly affected by the bremsstrahlung process
leading to a worse precision.
Therefore the analyses needed to perform checks of the precision in 
the specific topology under study. 

Since it was not possible to determine the production point of the tau pair,
the centre of the interaction region was used.  
An error was induced by this approximation 
due to both the size of the beams and the 
accuracy of the estimated position of the centre of the interaction region,
which was reconstructed from the
distribution of the primary vertices of hadronic Z decays.
In the vertical ($y$) direction the beams were very narrow and the main 
contribution
came from the uncertainty on the vertex reconstruction, which resulted in a 
typical precision of 10~$\mu$m in the production point position.
In the horizontal ($x$) direction, the uncertainty was dominated by the 
beam size and 
depended on the LEP operating conditions. The resulting uncertainty on the 
production point position ranged from about 90~$\mu$m in 1992 to 160~$\mu$m 
in 1995.

In addition to the above mentioned tracking detectors,
the identification of the $\tau$
decay products relied on electromagnetic calorimetry
for electron identification, and on hadron calorimetry and muon
chambers for muon identification. 

\begin{itemize}
\item
The barrel electromagnetic calorimeter was a High-density Projection
Chamber (HPC), covering the polar angle region from $43^{\circ}$ to 
$137^{\circ}$.          
It had a high granularity and provided nine layers of sampling
of shower energies.
\item
The Hadron Calorimeter (HCAL) was situated outside the magnet solenoid
and had a depth of 110 cm of iron. It was sensitive to
hadronic showers and minimum
ionizing particles and consisted of four layers 
with a granularity of $3.75^{\circ}$ in polar angle and
$2.96^{\circ}$ in azimuthal angle. 
\item
The barrel Muon  Chambers (MUB) consisted of  two layers of
drift chambers, the  first one situated outside 90  cm of iron
and  the   second  outside  the  hadron   calorimeter.   The
acceptance in  polar angle of  the outer layer was slightly
smaller  than the  other  barrel detectors  and covered  the
range   $|\cos\theta|\!<\!0.602$. The    polar   angle   range
$0.743\!<\!|\cos\theta|\!<\!0.940$ was covered by the forward Muon Chambers
(MUF). In 1994 a layer of Surround Muon Chambers (SMC), based on 
limited streamer tubes, was installed to fill the gap between the barrel 
and forward regions. 
\item
The TPC also provided up to 192 
ionisation measurements per charged particle track,
which were useful for electron/hadron separation.
\end{itemize}
 
Section~\ref{sec:vertex} describes the decay length analysis
applied to three-prong tau decays.
The two one-prong measurements, together with the correlations induced
from using overlapping data samples, are described in Section~\ref{sec:one}.
Section~\ref{sec:combi} presents the combined
result for the full 1991-1995 data sample. 
Conclusions are reported in Section~\ref{sec:summary}.

\section{The Decay Vertex Method }
\label{sec:vertex}
 
This method is an improved version of the one described in~\cite{DELPHI95}.
Decays of the Z into two taus were selected, where both taus 
decayed to three charged particles plus neutral particles (the 3v3 topology), or where one tau decay
contained three charged particles and the other decay one charged particle (the 3v1 topology). In the 3v1 topology only the three-prong decays were used. 

The event was divided into two hemispheres defined by the plane
perpendicular to the
thrust axis and passing through the centre of the interaction region.
The highest momentum charged particles in the two 
hemispheres were required to be separated by an angle of at least 2.9 and 
2.975~radians 
in the 3v1 and 3v3 topologies respectively. The tighter cut for the 3v3
topology reflects the increased background from hadronic decays of the Z.
For each tau decay, the three charged particle tracks had to be  
accurately reconstructed: 
each track had to be  
associated to at least two microvertex detector hits and all three tracks had
to be
consistent with coming from a common vertex with a $\chi^2$ probability 
greater than 0.5\%.
If the momentum of any charged particle 
in an event was greater than 35~GeV/$c$, 
there had to be no indication that this particle was either an electron or a muon.
This was necessary in order to eliminate four-fermion events with $\ee\ee$ and $\ee\mm$ 
final states. After this selection the only appreciable background 
consisted of $\ee\tt$ events and hadronic Z decays.

From simulation the background fractions, 
$b_f^{\msub{3v1}}$ and $b_f^{\msub{3v3}}$, of  four-fermion $\ee\tt$ events 
remaining in the 3v1 and 3v3 topologies were estimated to be 
$(0.15\pm0.03)$\%\ and $(0.5\pm0.2)$\% respectively.
These events had a decay length which was on average 25\%\ smaller than those
produced directly, due to the lower effective $\tau^+\tau^-$
centre-of-mass energy.  
This corresponded to a bias of $-0.1$ fs in the determination of the lifetime.

The background from hadronic Z decays was reduced by requiring at most six
electromagnetic neutral deposits in the event. By relaxing the cut on 
the opening angle and comparing the amount of additional events in data
and simulation, 
the hadronic contaminations 
in the 3v1 and 3v3 topologies,
$b_h^{\msub{3v1}}$ and $b_h^{\msub{3v3}}$ were estimated to be 
$(0.38\pm0.06)\%$ and $(0.8\pm0.2)\%$ respectively. 
These contributions were included in the lifetime fit with their errors
taken as systematic uncertainties.  
The hadronic background had no lifetime content and corresponded to a
bias of $-1.2\pm 0.2$~fs ($-2.2\pm 0.5$~fs) in the lifetime determination
in the 3v1 (3v3) topology.

Within the 1991--1995 data sample, 15427 and 2101 three-prong tau decays
 were selected in 
the 3v1 and 3v3 topologies respectively. The distance in the transverse plane 
between the reconstructed decay point of the tau and its production point,
taken as the centre of the beam crossing region, was calculated taking into
account the reconstruction uncertainty of the former and the size of the 
latter. 
This was
converted to a decay time $t$ and associated uncertainty $\sigma$ by 
dividing by $ \gamma\beta(\sqrt{s}) c\sin\theta$ where $\theta$ is the polar angle of the
thrust axis of the event, $c$ is the speed of light and
$\gamma\beta(\sqrt{s})=\sqrt{s/4m^2_\tau-1}$, with 
$\sqrt{s}$ the centre-of-mass energy of the collision, taken as twice the beam
energy with a correction for initial and final state radiation that was 
evaluated from the simulation to be between 
0.8\%\ and 1.4\%\ depending on the beam energy. The accuracy in the energy 
correction resulted in a  systematic uncertainty of 0.1~fs on the 
lifetime determination.

Simulation studies showed that the central value of the pull distribution
(defined as the reconstructed decay time
less the true decay time, divided
by the uncertainty) had a slight positive shift of $0.02\pm0.01$, leading to a
lifetime bias of $1.6\pm0.8$ fs.  This bias was subtracted from each event
with the uncertainty taken as a systematic uncertainty on the final lifetime 
determination.

The tau lifetime, $\tau_\tau$, was extracted from a maximum log-likelihood fit 
to the data using the function
\begin{equation}
L (\tau_\tau, \lambda_0^\msub{yr}, \lambda_1^\msub{yr}, \lambda_2^\msub{yr})
= 
\Sigma_i \log P\left( t_i|\sigma_i,\tau_\tau,
\lambda_0^{\msub{yr}}, \lambda_1^\msub{yr}, \lambda_2^\msub{yr}\right),
\quad \mbox{yr=1991--1995},
\end{equation}
where $t_i$ is the measured proper time for event $i$, after applying the 
correction for the biases resulting from reconstruction and alignment 
of the Microvertex Detector, $\sigma_i$ its 
uncertainty as computed by the reconstruction program and $\lambda^\msub{yr}$
are scaling factors as explained below. 

The probability density 
function $P$ is given by the convolution of the probability density 
function of the positions of decay vertices and the resolution function of the detector.
The sum runs over the full sample of events, spanning the five 
data-taking periods used in the analysis.

The probability density function of the decay vertices contains the dependence upon the lifetime:
\begin{equation} 
 f_\msub{vtx}(t|\tau_\tau) = 
   (1-b_f^\msub{topo}-b_h^\msub{topo}) E(t|\tau_\tau) 
 + b_f^\msub{topo} E(t|0.75\tau_\tau) 
 + b_h^\msub{topo}\delta(t),
 \quad \mbox{topo=3v1, 3v3,}
\end{equation}
where $E(x|\eta)$ denotes a normalized exponential distribution
with a decay constant $\eta$ and $\delta(t)$ is a Dirac delta function. 
The factor 0.75 in the four-fermion background 
term takes into account the reduced centre-of-mass energy for these events.

The lifetime component was convoluted with a resolution function $f_\msub{res}$ 
which,
according to the simulation, could be adequately parameterized as the sum of
three Gaussian distributions:
\begin{equation}
\label{eqn:res3}
f_\msub{res}(t|\sigma,f_2,f_3,k_1,k_2,k_3)=
(1-f_2-f_3)G(t|k_1\sigma) + f_2G(t|k_2\sigma) + f_{3}G(t|k_3\sigma),
\end{equation}
where $G(x|\eta)$ is a normalized Gaussian function centred at zero 
and with width $\eta$.
$f_2$ and $f_3$ are the fractions of the second and third Gaussian
functions, and $k_1\sigma,k_2\sigma,k_3\sigma$ are the widths.
The shape of the resolution function was taken from the simulation,
where the three Gaussian functions had widths of
$k_1=0.97,k_2=1.6,k_3=5.1$ times $\sigma$ and the fractional
contributions  of the second and third Gaussians were
$f_2=0.25$ and $f_3=0.007$ respectively.
However, there was some indication that the widths and proportions of
the Gaussian functions were slightly different in the real data and, moreover,
that they varied from year to year. 
Three scale factors for each year of data-taking,
$\lambda_0^\msub{yr}$, $\lambda_1^\msub{yr}$, $\lambda_2^\msub{yr}$ were
introduced 
in the likelihood function and these affect
the fraction, $f_2$, of the second Gaussian
and the widths of the first and second Gaussian functions, $k_1 , k_2$.

\begin{figure}

\epsfig{file=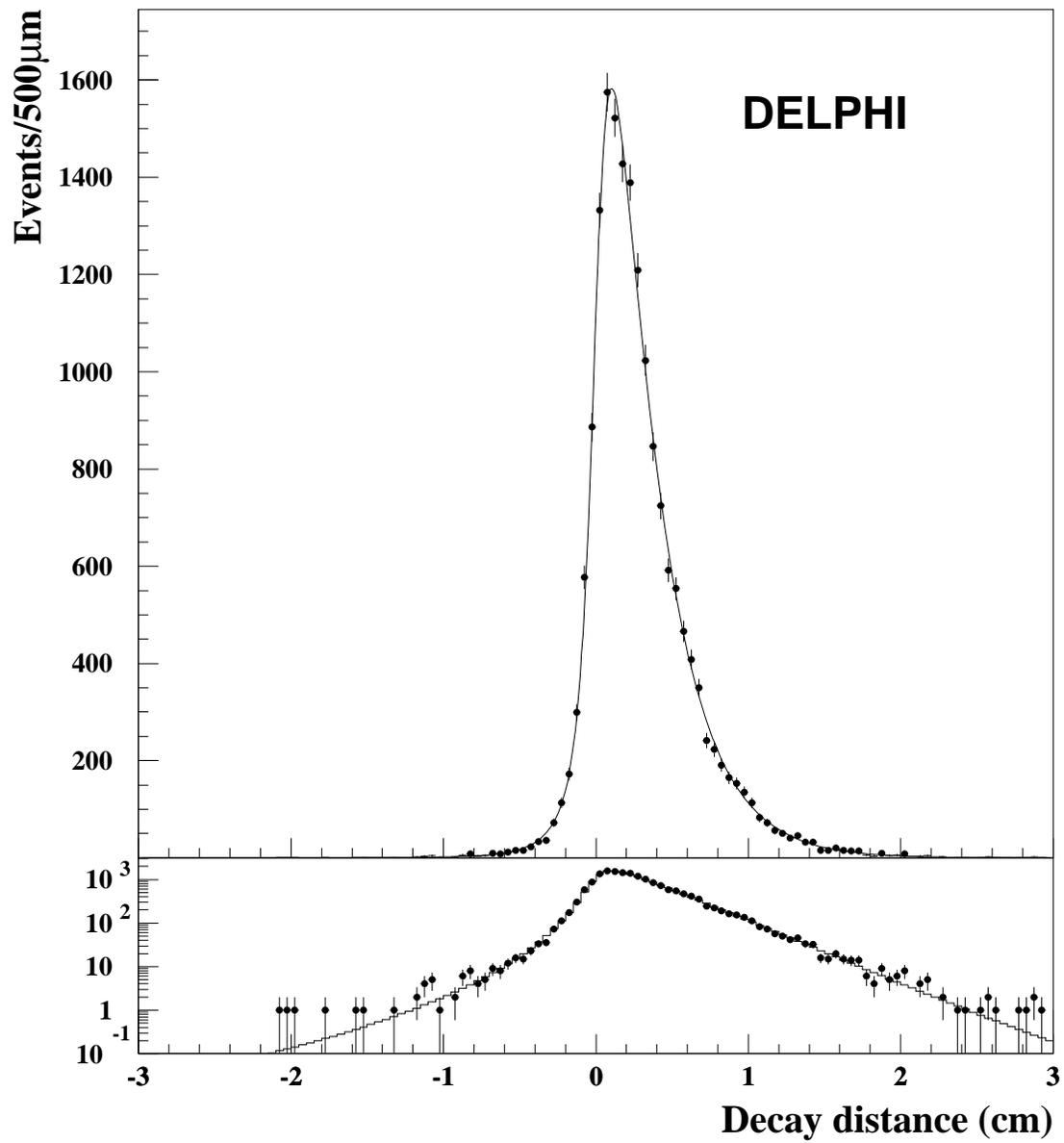,width=\textwidth}

\begin{center}
\end{center}
\caption{Decay length distribution for three-prong tau decays.
  The points are 1991-1995 data sample and the histogram is the result of the fit.}
\label{f:dlength}
\end{figure}

The complete expression for $P$ is given by
\begin{equation}
P\left( t|\sigma_i,\tau_\tau,
\lambda_0^{\msub{yr}}, \lambda_1^\msub{yr}, \lambda_2^\msub{yr}\right)=
f_\msub{vtx}(t|\tau_\tau)
\otimes
f_\msub{res}(t|\sigma_i,(\lambda_0^{\msub{yr}}f_2),f_3,
                        (\lambda_1^\msub{yr}k_1),
                        (\lambda_2^\msub{yr}k_2),k_3).
\label{eqn:full3p}
\end{equation}
The scale factors were fitted together with the lifetime and their values
are reported in Table~\ref{tab:result3}.
The tau lifetime was estimated to be $288.6\pm2.4$ fs, and
most of the scale factors were seen to be consistent with unity, confirming
that the shape and parameterization of the resolution function in
the simulation were reasonable.
The observed decay length distribution compared to the result of the maximum
likelihood fit is shown in Fig.~\ref{f:dlength}. 

\begin{table}
\vspace {0.3cm}
\begin{center}
\begin{tabular}{|l|c|c|c|} \hline
Year & \multicolumn{3}{c|}{Scale factors}    \\
\cline{2-4}
     & $\lambda_0$ & $\lambda_1$ & $\lambda_2$ \\ 
\hline
1991 &$1.13 \pm 0.29$ & $1.30 \pm 0.25$ & $1.66 \pm 0.62$ \\
1992 &$0.79 \pm 0.51$ & $0.87 \pm 0.21$ & $0.86 \pm 0.17$ \\
1993 &$1.03 \pm 0.18$ & $0.86 \pm 0.10$ & $1.08 \pm 0.17$ \\
1994 &$1.15 \pm 0.07$ & $0.99 \pm 0.05$ & $1.43 \pm 0.18$ \\
1995 &$1.06 \pm 0.17$ & $0.93 \pm 0.10$ & $1.21 \pm 0.22$ \\
\hline
\end{tabular}
\caption{Fitted values for the resolution function parameters 
in (\ref{eqn:full3p}).}
\label{tab:result3}
\end{center}
\end{table}

\begin{table}
\vspace {0.3cm}
\begin{center}
\begin{tabular}{|l|c|c|} \hline
Year & Alignment correction & Contr. to syst. uncertainty   \\
\hline
1991  &$83\pm45\ \mu$m &$\pm 0.23$ fs\\
1992  &$-2\pm19\ \mu$m &$\pm 0.44$ fs\\
1993  &$\ 0\pm17\ \mu$m&$\pm 0.42$ fs\\
1994  &$21\pm13\ \mu$m &$\pm 0.57$ fs\\
1995  &$32\pm22\ \mu$m &$\pm 0.47$ fs\\
\hline
{\bf Total} & & $\pm 0.98$ fs\\ 
\hline
\end{tabular}
\caption{Alignment corrections from hadronic data and their 
contributions to the total systematic uncertainty.}
\label{tab:align3}
\end{center}
\end{table}

%
%

A number of consistency checks were performed.
Firstly, the maximum likelihood fit was repeated holding
all the scale factors fixed to unity (i.e. taking the resolution function
from the simulation) and this gave a value of $\tau_\tau=289.1$ fs.
Secondly, the resolution function was taken from the simulation with a 
single scale factor allowed to multiply $k_1,k_2$ and $k_3$.  This
gave a value of $\tau_\tau=289.0$ fs.
Thirdly, starting with the simulation resolution function, five
scale factors (one for each year) were allowed to multiply $f_3$,
in order to gauge the effect of the smallest but broadest Gaussian.
This gave a value of $\tau_\tau=288.0$ fs.
Fourthly, a weighted mean\footnote{The weighted mean is defined as
$\sum_i (t_i/w_i)/\sum(1/w_i)$ where 
$w_i=(\sigma_i^{RMS})^2+\tau_\tau^2$ and $\sigma_i^{RMS}$ is the RMS of
the resolution function for a given $\sigma_i$.}
of the data, corrected for background
biases was computed. This test is insensitive to the exact shape of the 
distribution and it gave a value of $\tau_\tau=288.0$ fs.

The largest systematic uncertainty came from the accuracy of the alignment 
of the Microvertex Detector. This was calculated year-by-year by taking a 
sample of hadronic decays of the Z with three tracks in one hemisphere and
more than three tracks in the other. In such topologies,
the momentum and invariant mass distributions of the three tracks were 
similar to those in tau decays. 
A vertex was formed from the three tau-like tracks and
the distribution for the decay distance was compared to a simulation of 
hadronic decays of the Z. 
Possible indications for shifts in the reconstructed position 
were observed and are summarized in Table~\ref{tab:align3}.
These corrections were applied to the reconstructed decay distances before the
maximum likelihood fit was made.
The related systematic uncertainty was calculated for each year by changing the 
alignment correction for that year by its uncertainty and repeating the complete 
fit using all years. The alignment corrections were uncorrelated between years
and so their systematic uncertainties were added in quadrature to give $\pm0.98$~fs. 

\begin{table}
\begin{center}
\begin{tabular}{|l|r|}
\hline
Fitted lifetime & $ 288.6\pm 2.4 $ fs  \\
\hline
Background           & $\pm0.2$ fs \\
Radiative Energy Loss& $\pm0.1$ fs \\
Reconstruction Bias  & $\pm0.8$ fs \\
Alignment            & $\pm1.0$ fs \\
\hline
{\bf Total}          &  $\mathbf 288.6\pm 2.4\pm 1.3 $ {\bf fs}\\
\hline
\end{tabular} 
\caption{Summary of fit result and systematic uncertainties for the 
3-prong topology. }
\label{tab:sys3} 
\end{center}
\end{table}

A summary of the systematic uncertainties due to each source is given 
in Table~\ref{tab:sys3}.
The tau lifetime from three-prong decays of the tau lepton was thus
measured to be
$$
\tau_\tau \mbox{(3 prong)} = 288.6\pm 2.4_{stat} \pm 1.3_{sys} \mbox{ fs}.
$$
This result supersedes previously published three-prong DELPHI results.

\section {One-Prong Lifetime Measurements}
\label{sec:one}
  
The lifetime information from one-prong tau decays was obtained
by measuring the impact parameters of the charged decay products.
In the case of perfect knowledge
of the track
parameters and of the production point,
the impact parameter in the $R\phi$ plane is given by:
\begin{equation}
d = L \sin\theta_{\tau}  \sin( \phi - \phi_{\tau} ),
\label{eqn:impact_generic}
\end{equation}
where $L$ is the decay length, $\phi_{\tau}$ the azimuthal direction of
the decaying object, $\phi$ the track's azimuth and
$\theta_{\tau}$ the polar angle of the decaying object.
Impact parameters, signed according to the geometrical definition 
given in Section~\ref{s:intro}, were used in the
calculation of the resolution functions, as well as in the extraction
of the tau lifetime. 

The tau lifetime was extracted from events in which both taus decayed
to a single charged particle using two methods.
The first method used the impact parameter difference (IPD) which
represents an improvement over the single hemisphere impact
parameter lifetime determination (see for example~\cite{DELPHI91})
by reducing the dependence of the measurement on the unknown tau-decay angle.
In the IPD method, knowledge of the
tau-pair production point was limited by the size of the interaction region,
whose dimensions are larger than the track extrapolation precision
of the detector. To avoid this limitation, the second method used the
track pair miss distance (MD). In the MD method 
the two impact parameters in a
$\tau^+\tau^-$ event were summed 
so that the dependence on the production
point inside the interaction region cancelled to first order.
This second method was sensitive to the knowledge of the
resolution function, which is described in detail in
Section~\ref{sec:miss}.
 
The $\tau^{+}\tau^{-}$ data were selected in the same way as those
used for the tau polarisation measurement \cite{polpap}. The event 
was divided into two hemispheres by a plane perpendicular
to the thrust axis and events were required to have:
\begin{itemize}
\item
the highest momentum charged particle in at least one of the two hemispheres
with $|\cos\theta|<0.732$;
\item
total charged particle multiplicity less than 6;
\item 
isolation angle between the highest momentum particles in the two hemispheres
greater than $160^\circ$;
\item
the highest momentum particles in the two hemispheres passing at less than 4.5
cm in $z$ and less than 1.5 cm in the $R\phi$ plane from the center of the
interaction region and the difference in $z$ of the points of closest approach
should be less than 3 cm;
\item
total energy of the event greater than 8 GeV and total transverse momentum 
greater than 0.4 GeV/c;
\item
acollinearity of the two highest momentum tracks in each hemisphere greater than
$0.5^\circ$;
\item 
$P_{\mbox{\scriptsize rad}}$ less than the beam momentum and 
$E_{\mbox{\scriptsize rad}}$ less than the beam energy. 
\end{itemize}
In the above $P_{\mbox{\scriptsize rad}}=(|\vec{p}_1|^2+|\vec{p}_2|^2)^{1/2}$
and $\vec{p}_{1}$ and $\vec{p}_{2}$ are the charged-particle momenta and
$E_{\mbox{\scriptsize rad}}=(E_1^2+E_2^2)^{1/2}$. The variables
$E_1$ and $E_2$ are the total electromagnetic
energies deposited in cones of half-angle $30^\circ$
about the charged-particle momentum vectors $\vec{p}_1$ and
$\vec{p}_2$ respectively.

In addition there must be only one
charged particle track per hemisphere with hits in the 
Microvertex Detector.
Both tracks had to satisfy the following requirements:
a momentum transverse to the beam axis greater than 1 GeV/$c$,
associated $R\phi$ hits 
in at least two VD layers, at least 11 associated points in
the TPC and a
$\chi^2$ probability about 0.1\% for the additional $\chi^2$ when all 
of the VD hits are added to the track fitted to the TPC segments.
 

After the application of the above criteria, the remaining
background 
came essentially from
$\ee$, $\mm$ and  $\gamma\gamma\rightarrow\ell\ell$ events. 
Further selections were applied depending on the result 
of the lepton
identification described in Section~\ref{s:leptid}.
To suppress $\mu^+\mu^-$ events, if both charged particles were identified as
muons, $P_{\mbox{\scriptsize rad}}<35$~GeV/c was required.
To suppress e$^+$e$^-$ events,
if one of the particles was identified as an electron and the other was
not a muon, the requirements
$P_{\mbox{\scriptsize rad}}<35$~GeV/c and
$E_{\mbox{\scriptsize rad}}<30$~GeV were imposed.
Finally two-photon events were suppressed requiring
$P_{\mbox{\scriptsize rad}}>11$~GeV/c 
or $E_{\mbox{\scriptsize rad}}>8$~GeV,
if both charged particles were identified as muons or electrons.

These criteria selected 17366 and 8670 events in the 1994 and 1995 data
samples respectively.
The residual background in the sample was obtained from simulated 
background events and amounted to
$(0.58\pm 0.05)\%$ from dilepton events and to $(0.31\pm 0.03)\%$
and $(0.46\pm0.04)\%$ in 1994 and 1995 respectively from $\gamma\gamma$
events. The quoted uncertainties were derived from the available statistics of
 simulated events.

\subsection{Lepton Identification}
\label{s:leptid}

Particle identification had an influence on different aspects of this
analysis, namely background suppression and, for the impact parameter
sum method, also in the physics and resolution function 
parameterization (see Section~\ref{sec:miss}). 
A neural network was implemented 
to improve the particle identification, using the same 
variables as for the particle identification in $\tau$ decay applied in \cite{Mogen}: 
%
\begin{itemize}
\item
the neutral electromagnetic energy measured by the HPC 
  in a cone of half-angle 
  $19^{\circ}$ about the track;   
\item
the number of hits associated in the first HCAL layer;
\item
the number of hits associated in the last HCAL layer;
\item
the average energy deposited in a HCAL layer;
\item
the number of hits associated in the muon chambers;
\item
the pull functions of the measured energy loss 
$dE/dx$ compared with the one expected from an electron or
from a pion\footnote{The pull function  is 
$\frac{(dE/dx)_{\msub{meas}}-(dE/dx)_{e/\pi}}{\sigma_{dE/dx}} $,
  where $(dE/dx)_{\msub{meas}}$ is the energy loss measured by the TPC, 
  $(dE/dx)_{e/\pi}$ is the one expected from an electron (or a pion) and $\sigma_{dE/dx}$ is
  the energy loss reconstruction uncertainty.};
\item
the $E/p$ of the particle, where $p$ is the momentum measured from the curvature of the track
  and $E$ is its associated electromagnetic energy;
\item
the $\chi^{2}$ probability for
    $dz=z_{\msub{HPC}}-z_{\msub{extr}}$, where $z_{\msub{HPC}}$
  is the $z$ of the shower associated to the track,
  measured by the HPC and $z_{\msub{extr}}$ 
  the one extrapolated from the TPC to the HPC.
\end{itemize}
A different neural network was implemented for the small fraction ($2.5 \%$) of
events with no reliable $dE/dx$ measurements.
Both neural networks were three-layer feed-forward neural networks,
with only 
two output nodes, one for electrons and one for 
muons. Hadrons were classified as all 
charged particles not identified as either electron or muon.
 
The neural network performance was tested with  simulated samples different 
from those used 
for the training. For a simulated sample corresponding to 1995 data, an electron
identification efficiency and purity of $(95.6\pm0.1)\%$ 
and $(91.8\pm0.1)\%$ respectively were obtained, while for muons 
the corresponding numbers were $(96.7\pm0.1)\%$ 
and $(95.0\pm0.1)\%$. 



\subsection{The Impact Parameter Difference Method}
\label{sec:ipd}

At LEP, the taus were produced with high boost and in back-to-back pairs
which allowed some simplifications~\cite{ALEPH} in evaluating
equation (\ref{eqn:impact_generic}).
The boost allows the sine of the decay angle to be approximated by
$\sin(\phi-\phi_\tau)\approx \phi-\phi_\tau$, while the collinearity
allows the substitutions $\sin\theta_{\tau^-}=\sin\theta_{\tau^+}$ and
$\phi_{\tau^-}=\phi_{\tau^+}\pm\pi$. 
Therefore, considering both tracks in the event,
\begin{equation}
d_+ - d_- = \sin\theta_{\tau^+} \left( L_{+}\phi_+ - L_{-}\phi_-
                                  -(L_{+}-L_{-})\phi_{\tau^+}
                                  \pm L_{-}\pi \right)
\label{eqn:ipd}
\end{equation}
where the variables are the same as in equation (\ref{eqn:impact_generic}), 
with the additional subscript to indicate the charge of the tau in the pair.
The sign of $\pi$ is chosen to normalize the acoplanarity of the
two observed tracks, $\Delta\phi=\phi_+-\phi_-\pm\pi$, to be in 
the  $\pm\pi/2$ range.
An average over the decay lengths shows that the 
average difference of impact parameters is related to the
projected acoplanarity, $\sin\theta_\tau \Delta\phi$, of the decay products~\cite{ALEPH}:
\begin{equation}
\langle d_+ - d_- \rangle = \gamma\beta(\sqrt{s}) c\tau_\tau 
         \sin\theta_\tau \Delta\phi.
\label{eqn:ipdaverage}
\end{equation}

Equation (\ref{eqn:ipdaverage}) 
shows that the lifetime information can be extracted from the measurement
of the directions of the outcoming decay products,
independently of the poorly known production angle $\phi_\tau$.
The uncertainty on the impact parameter difference is, however, affected 
strongly by the uncertainty on the production point. Moreover the small
angle and tau-pair collinearity are just approximations that cease to 
be valid for substantial decay angles and for hard
initial-state radiation. Therefore deviations from the 
linear behaviour of equation  (\ref{eqn:ipdaverage}) are expected at large 
values of acoplanarity.

%
%
As discussed in a previous publication \cite{DELPHI95}, the average decay 
length 
$\langle L\rangle = \gamma\beta(M_Z) c \tau_\tau$ was determined from an event 
weighted $\chi^2$ fit of a straight line to the impact parameter 
difference as a function of the projected acoplanarity:
\begin{equation}
\langle L \rangle = 
     \frac{\sum w \sum wxy - \sum wx \sum wy}{\sum w \sum wx^2 - (\sum wx)^2},
\end{equation} 
where the sum extends over the selected events,
\begin{eqnarray}
y & = & d_+-d_-, \\
x & = & 
  \left( \frac{\gamma\beta({\sqrt s})}{\gamma\beta(M_Z)}\right)
  \sin\theta_\msub{thrust} ( \phi_+ - \phi_-).
\end{eqnarray}
The weight $w$ given to each single measurement depends upon several 
variables: 
\begin{equation}
w   =   \left[
        \sigma^2_\msub{phys} 
      + \sigma^2_\msub{x,beam}(\sin\phi_+ - \sin\phi_-)^2
      + \sigma^2_\msub{y,beam}(\cos\phi_+ - \cos\phi_-)^2
      + \sigma^2_{d_+}
      + \sigma^2_{d_-}
      \right]^{-1},
\end{equation}
where 
\begin{itemize}
\item $\sigma_\msub{phys}$ 
is the {r.m.s.} of the distribution of the impact
parameter difference due to the lifetime spread;
\item $\sigma_\msub{x,beam}$ and $\sigma_\msub{y,beam}$
are the uncertainties in the production point due to the beam size and 
the knowledge of beam position
 in the $x$ and $y$ DELPHI coordinates (see Section~\ref{s:intro}); 
\item $\sigma_{d_\pm}$
is the impact parameter uncertainty from the DELPHI track fit.
\end{itemize}


An outlier rejection 
was performed discarding the events with the highest
significance of the residual $|\sqrt{w}(y-\langle L\rangle x)|$.
This procedure, designed to reduce the statistical fluctuations in the
final result due to badly reconstructed events, 
was used to reject at most 1\% of the sample.

\begin{figure}
\begin{center}
\epsfig{file=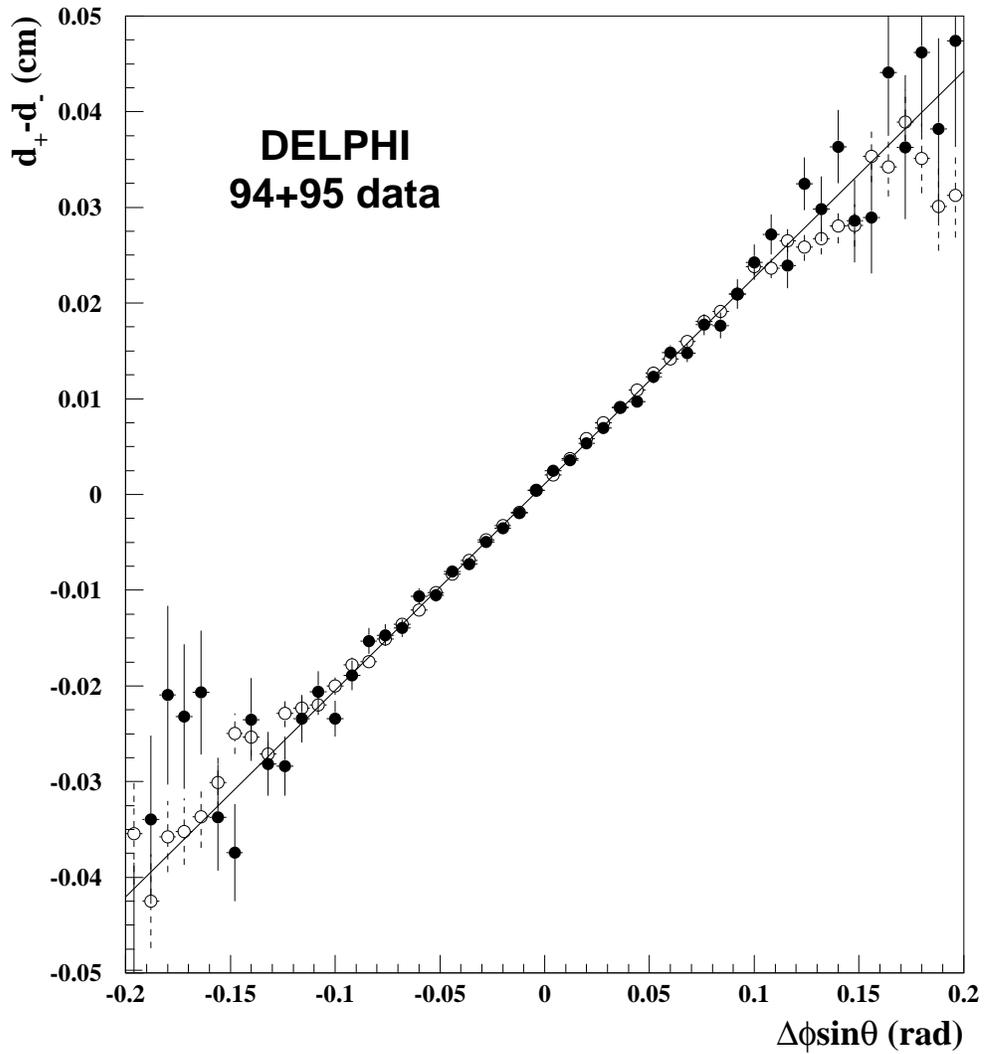,width=\textwidth}
\end{center}
\caption{Distribution of the difference of impact parameters versus
  the projected acoplanarity for the joint '94-'95 data sample ($\bullet$)
  and simulation ($\circ$); the straight line is the best weighted 
  $\chi^2$ fit on all the 94-95 events.
}
\label{f:ipd9495}
\end{figure}

The impact parameter difference versus the projected acoplanarity 
 is shown
in Fig.~\ref{f:ipd9495}, together with the result of the fit.
The results of the fits to 1994 and 1995 data were:
\begin{eqnarray}
\langle L \rangle _{94} & = & 2161\pm 33 \mbox{\ $\mu$m}, \nonumber \\
\langle L \rangle _{95} & = & 2150\pm 51 \mbox{\ $\mu$m}, \nonumber
\end{eqnarray}
where only the statistical uncertainties are shown.
The $\chi^2/\mbox{DOF}$ of the fits were respectively  16279/16288 and 7544/8098
for 1994 and 1995 data, the ratio being lower than unity as expected 
due to the outlier rejection.

The IPD method suffered from several biases
which had their origin in the assumptions and approximations of the
method and in the need to trim the tails of the significance
distribution. Another source of bias was the background contamination. 

The assumption that collinear tau pairs had been produced with the
full centre-of-mass energy led to a bias of $-26.9\pm1.3$~$\mu$m which
was evaluated from the simulation of initial-state radiation.

\begin{figure}
\begin{center}
\includegraphics[width=\textwidth]{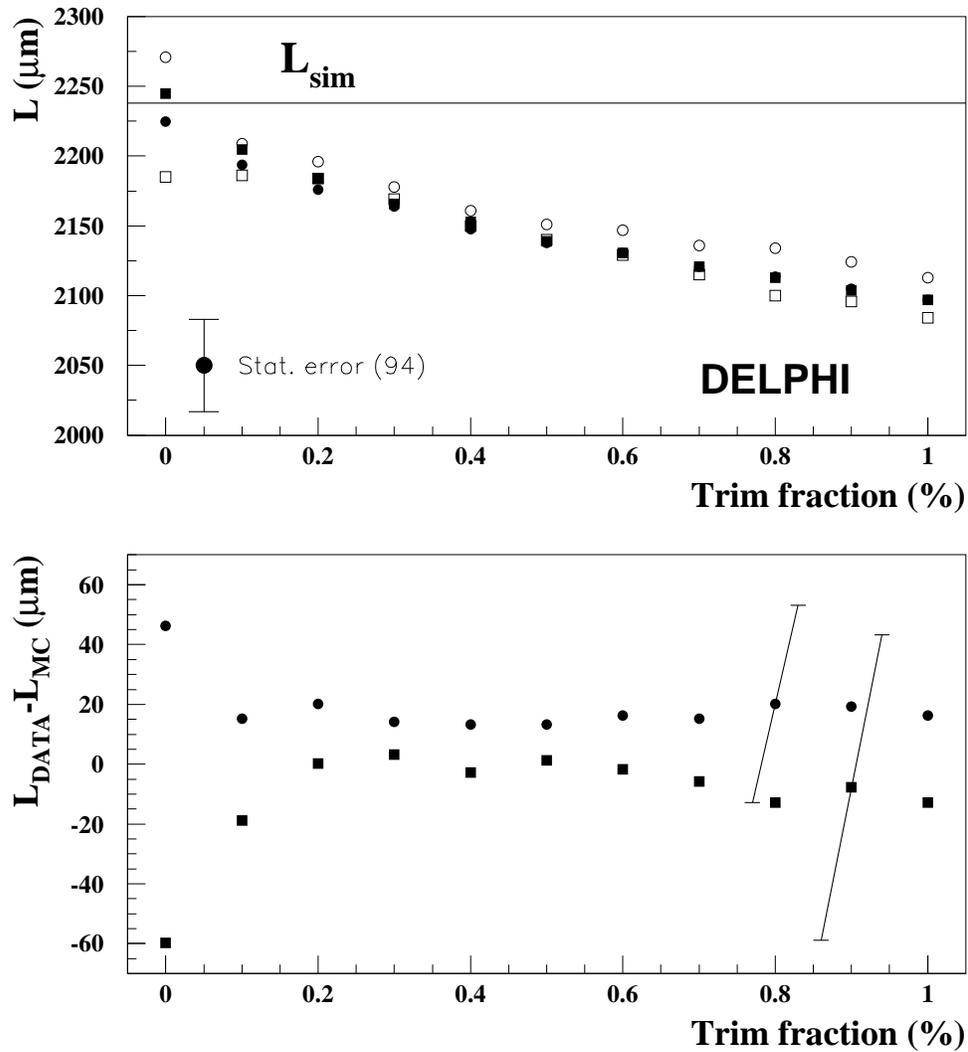}
\end{center}
\caption{
Upper panel: fitted decay length versus trim fraction: 
'94 simulation ($\bullet$), '94 data ($\circ$), 
'95 simulation ($\blacksquare$) and  '95 data ($\square$).
Lower panel: the difference
  between data and simulation for '94 ($\bullet$) and '95 ($\blacksquare$). 
  Statistical uncertainties 
  of the fit are added for comparison on two data points. 
  The simulated decay length is $L_\msub{sim}=2237$~$\mu$m.}
\label{f:trimdep}
\end{figure}

The above mentioned outlier rejection, 
performed by trimming of the residuals,
introduced an additional bias since the
asymmetric exponential tail due to the lifetime distribution was
preferentially cut. The effect was significant and the
induced shift was derived from simulated event studies.
A check was made comparing the behaviour of 
the decay length fitted on the
data with the expectation from the simulation. 
The fitted decay lengths as a function of the trim fraction are
displayed in Fig.~\ref{f:trimdep}. There was good agreement between
data and simulation, 
providing a lifetime determination that is 
stable with respect to changes in the trim fraction.
A trimming point of 0.4\%
was chosen as in previous publications, corresponding to
a bias of $-43.4\pm 7.9$ $\mu$m and $-40.3\pm 6.8$ $\mu$m for 1994 and
1995 respectively. The uncertainty quoted is the one due to simulation 
statistics.  
In order to take into account the uncertainty due to the modeling 
of the trim dependence in the simulation, a further contribution to the
systematic uncertainty was added.
It was
estimated as the maximum difference between the lifetime evaluated at
the chosen trim point and all the other evaluations in the interval [0.1;1.0]\%.
This amounted to 9 $\mu$m and 10 $\mu$m for 1994 and 1995 data
respectively.

As an additional systematic check, 
the point at which the line intercepted the $y$-axis was
determined. This was expected to be different from zero, due to the
correlation between the energy loss (particularly relevant for electrons) 
and the shift in the reconstructed impact parameter. 
This effect was checked using the
lepton identification algorithm described in Section~\ref{s:leptid}.
Combining the 1994 and 1995 samples,
the offsets were $23.0\pm 2.7$ $\mu$m for events
containing at least one identified electron and $3.9\pm 1.4$ $\mu$m
for events with no identified electrons. These results agreed well with the
expectations from the simulation of 
$20.6\pm 0.7$ $\mu$m and $5.4\pm 0.6$ $\mu$m 
respectively.

The bias induced by the background was estimated by
adding to a simulated sample 50 samples of simulated
background events which had passed the
selection criteria. The average bias was $-11.4\pm 4.9$ $\mu$m
for 1994 and $-17.5\pm 7.4$ $\mu$m for 1995, where
the systematic uncertainty was
the r.m.s. of the biases calculated in each of the 50 samples.
Additional systematic uncertainties were due 
to the uncertainty on the resolution function (3.8 $\mu$m)
and to the vertex detector 
alignment.
The latter was checked by computing the lifetime using a vertex detector 
geometrical description simulating the alignment uncertainties.
Most parameters are well constrained by the alignement procedure and
provide negligible variation in the lifetime determination. 
Only the less constrained deformation, a coherent radial variation  
of the silicon ladders, provided a visible shift in the reconstructed
values, which amounted to 3.1 $\mu$m, for a radial movement of
20 $\mu$m. These results are in qualitative agreement with what obtained 
from a simplified simulation in section 5.3 of \cite{Wasserbaech}.

%

\begin{table}
\begin{center}
\begin{tabular}{|l|r|r|r|}
\hline
                &     1994      &       1995      & Correlation \\ 
                & [fs]          &  [fs]           & coefficient  \\  \hline
Fit result      & $281.1\pm 4.3$ & $279.7\pm 6.7$ &     0   \\ \hline
Syst. sources:  &                &                &         \\
Method bias     & $ +3.5\pm 0.2$ & $ +3.5\pm 0.2$ &     1   \\
Trim            & $ +5.6\pm 1.0$ & $ +5.2\pm 0.9$ &     0   \\
Trim data/MC agreement
                & $     \pm 1.2$ & $     \pm 1.3$ &     0   \\
Background      & $ +1.5\pm 0.6$ & $ +2.3\pm 1.0$ &     0.45   \\
Alignment       & $     \pm 0.4$ & $     \pm 0.4$ &     0   \\
Resolution      & $     \pm 0.5$ & $     \pm 0.5$ &     1   \\
\hline
Result          & $291.7\pm 4.3\pm 1.8$ & $290.7\pm 6.7 \pm 2.0$ &
                0.02 \\
\hline
{\bf Average 94+95}&\multicolumn{3}{c|}{$\mathbf 291.4\pm 3.6\pm 1.5$} \\
\hline
%
\end{tabular} 
\caption{Summary of results, correction for biases, systematic uncertainties 
and combination of the Impact Parameter Difference measurements.  }
\label{t:IPDsyst}
\end{center}
\end{table}

The uncertainty due to the resolution function was
considered to be correlated between the two years and a correlation of 0.45
was calculated for the uncertainty due to the background.
All the other systematic uncertainties were uncorrelated, as they came from
calibrations which were computed with different data sets for the
two samples.
A summary of all systematic uncertainties is given in Table~\ref{t:IPDsyst}.
The lifetimes were measured to be:
\begin{eqnarray}
\tau_\tau\mbox{(IPD, 94)} &=&
     291.7 \pm 4.3_{stat}\pm 1.8_{sys} \mbox{\ fs}, \nonumber \\
\tau_\tau\mbox{(IPD, 95)} &=&
     290.7 \pm 6.7_{stat}\pm 2.0_{sys} \mbox{\ fs}, \nonumber \\
\tau_\tau\mbox{(IPD, 94+95)} &=&
     291.4 \pm 3.6_{stat}\pm 1.5_{sys} \mbox{\ fs}. \nonumber 
\end{eqnarray}

\subsection{The Miss Distance Method}
\label{sec:miss}
 
%
%
%

At LEP, 
the algebraic sum of the impact parameters, $\delta=d_++d_-$, named the ``miss 
distance'', was strongly correlated to the separation of the two tracks at 
the production point. The width of the miss distance distribution for
one-prong versus one-prong tau decays depends on the value of the 
lifetime. 
This was measured by an unbinned maximum likelihood
fit to the observed distribution. 
The probability density function was
given by the convolution of a physics function and a 
  resolution function.

%
%
The physics function was given by the 
distribution of miss distances expected 
from the decay length of $\tau$'s. This was built from the convolution 
of the impact parameter distribution of tracks originating from 
$\tau^+$ decays with that of tracks originating from $\tau^-$ decays.

The dependence of the impact parameter distribution upon 
the track momentum, the decay kinematics and the 
$\tau$ helicity was also considered.
The differences in the distribution for leptons and hadrons depend on 
the different $\tau$ decay kinematics in one-prong topologies; 
while leptonic tau decays have three final-state particles, hadronic
decays have two or more depending on the number of neutral particles
present in the decay.
These differences are important for momenta less than 20 GeV/$c$.

The shape of the single impact parameter distribution, as a function of 
decay dynamics and kinematics, was determined on a sample of 185842 
simulated events which passed all the selection and quality cuts; 
it was 
parameterized as a linear combination of three exponentials.

To obtain the miss distance, the convolution of the two
functions describing the single impact parameter distribution,
$f_{IP\: +}(d_{+})$ for the $\tau^{+}$  and
$f_{IP\: -}(d_{-})$ for the~$\tau^{-}$, was calculated.  Since 
the helicities were not known for a single event, the function
was calculated as the sum of the physics function for positive helicity
events and the one for negative helicity events, mixed according to the $\tau$
polarisation ~\cite{polpap2}. 

%

To compute the lifetime, the physics function was convoluted with
the experimental uncertainties on the reconstruction of the miss
distance.
As the lifetime information was only in the width of the 
miss distance distribution,
and not in its average value, the increase in width induced by the 
reconstruction uncertainties had to be 
evaluated with high precision. In particular
it was essential to have a good modeling of the tails due to
scattering of the particles through the apparatus.

\begin{figure}
\begin{center}
\includegraphics[width=\textwidth]{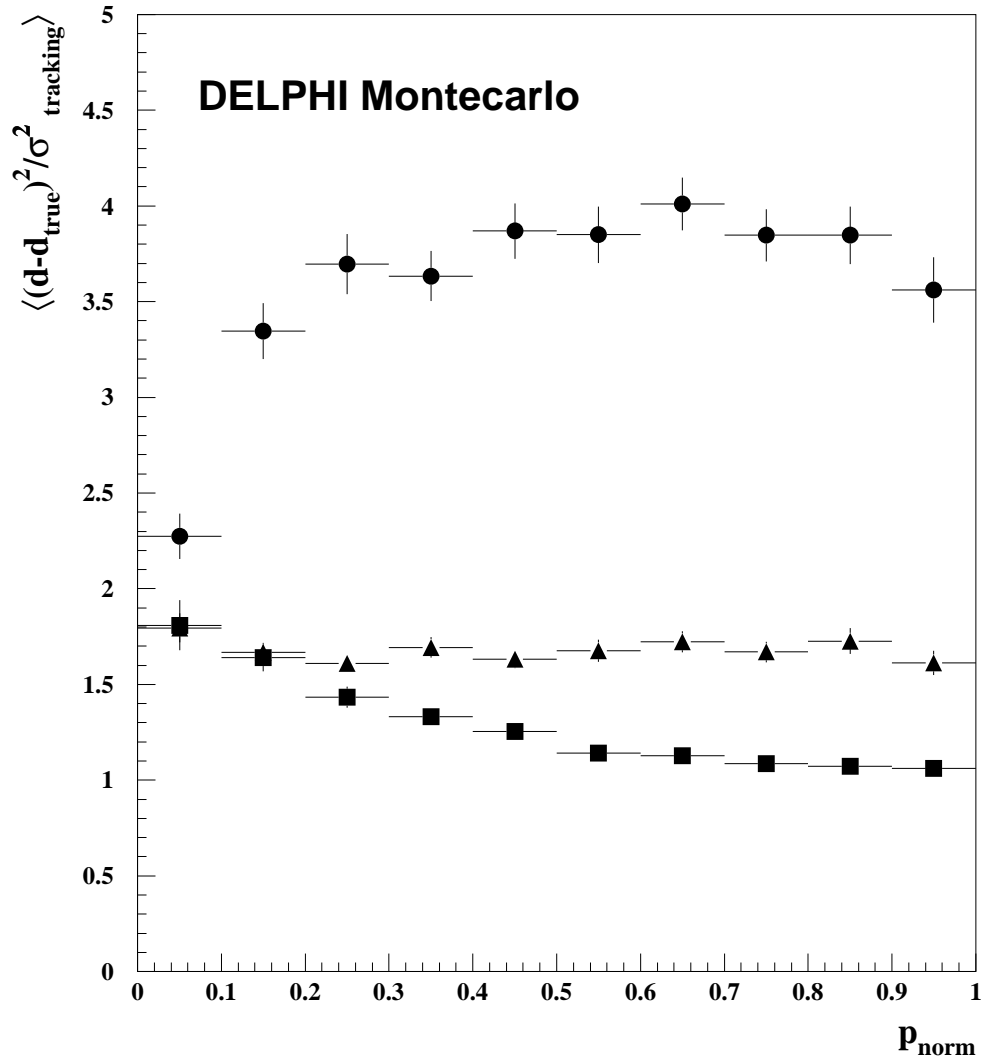}
\end{center}
\caption{Average value 
of the variance of the pull function of the reconstructed
impact parameter for simulated $\tau$ decays, 
as a function of the normalized momentum (see text) for 
hadrons ($\blacktriangle$), muons ($\blacksquare$) and electrons ($\bullet$).
By construction the lifetime information is removed from this pull.}
\label{f:pull1}
\end{figure}

The resolution was determined for the single impact parameter and then
convoluted to provide the miss distance. 
Hadrons and leptons required different resolution functions, as can be seen
in Fig.~\ref{f:pull1}, where the 
variance of the pull distribution of the reconstructed impact parameter 
is plotted against $p_\msub{norm}$,  for the simulated tau sample.
The pull is defined as the difference between the 
generated and reconstructed impact parameter divided by 
the tracking uncertainty given by the reconstruction program. 
The variable $p_\msub{norm}$ is defined as
\begin{equation}
p_\msub{norm} = \frac{p}{E_\msub{beam}}
                \left( 2-  \frac{p}{E_\msub{beam}} \right),
\end{equation}
where $p$ is the particle momentum.
This variable was chosen because its distribution 
is almost flat for tau decays.

Fig.~\ref{f:pull1} shows that
the average value of the variance of the pull is different from unity.
This was due to the presence of tails in the impact parameter distribution. 
At low momenta all types of particles had similar precision 
since the dominant effect was multiple Coulomb scattering
while at high momenta the 
different interactions of the particles show up in a difference in resolution.
The momentum dependence of the variance of the impact parameter pull for
hadrons was much less than that for muons and electrons.  This was a
consequence of the tuning procedure used to obtain the tracking uncertainties,
which was based largely on (both simulated and real data) samples of hadrons in
multi-hadronic Z decays.

To have an acceptable description of the resolution function, 
a three-Gaussian parameterization,
\begin{equation}
f_\msub{res}(d|p,\theta) = 
\sum_{i=1}^{3} f_i(p,\theta)G\left( d|\sigma_i(p,\theta)\right), 
\end{equation}
was used. Both the relative fractions $f_i$ and widths $\sigma_i$ 
of the Gaussians were dependent on the momentum and polar-angle,
as explained below.

\begin{figure}
\begin{center}
\epsfig{file=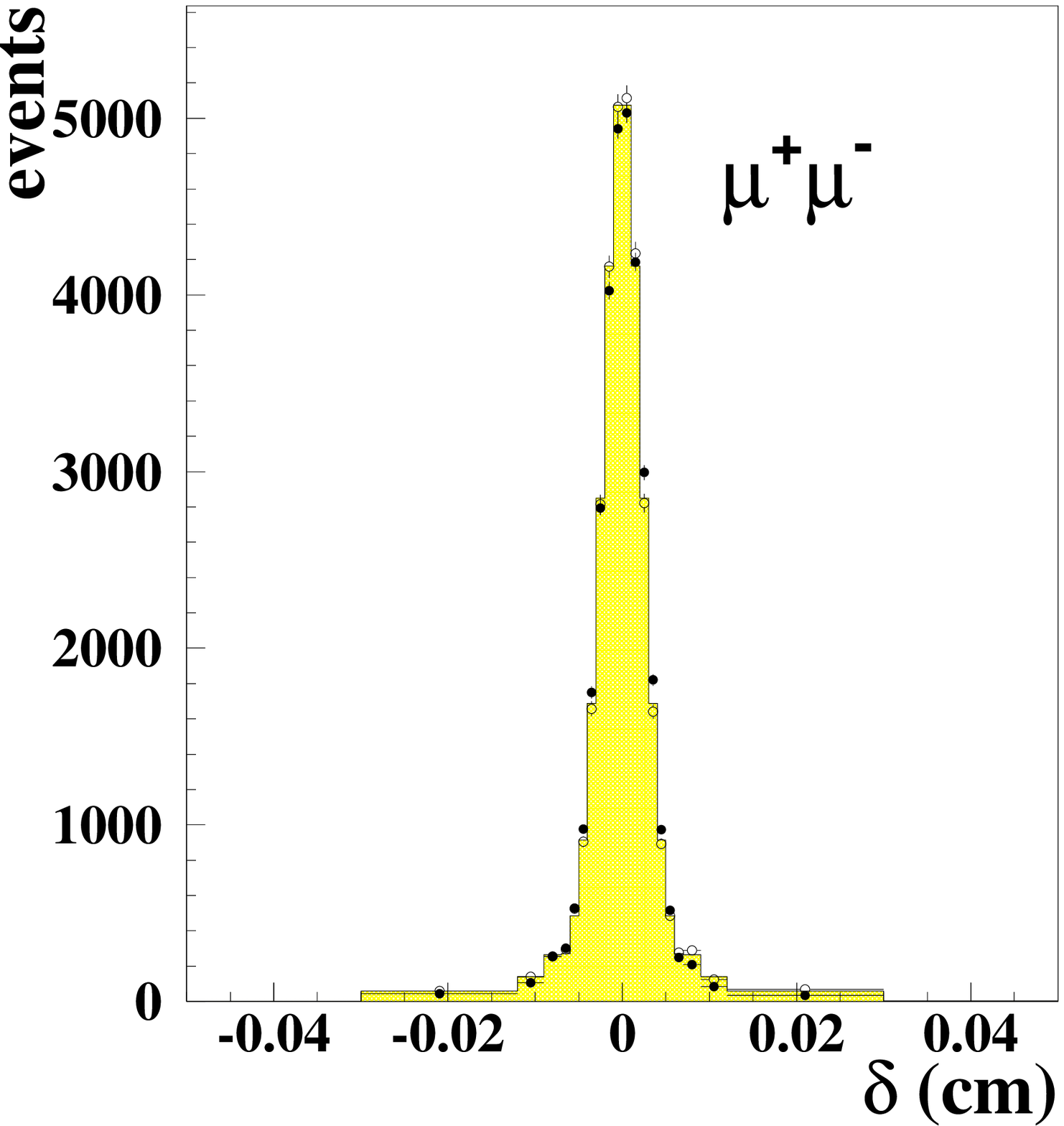,width=0.45\textwidth}
\epsfig{file=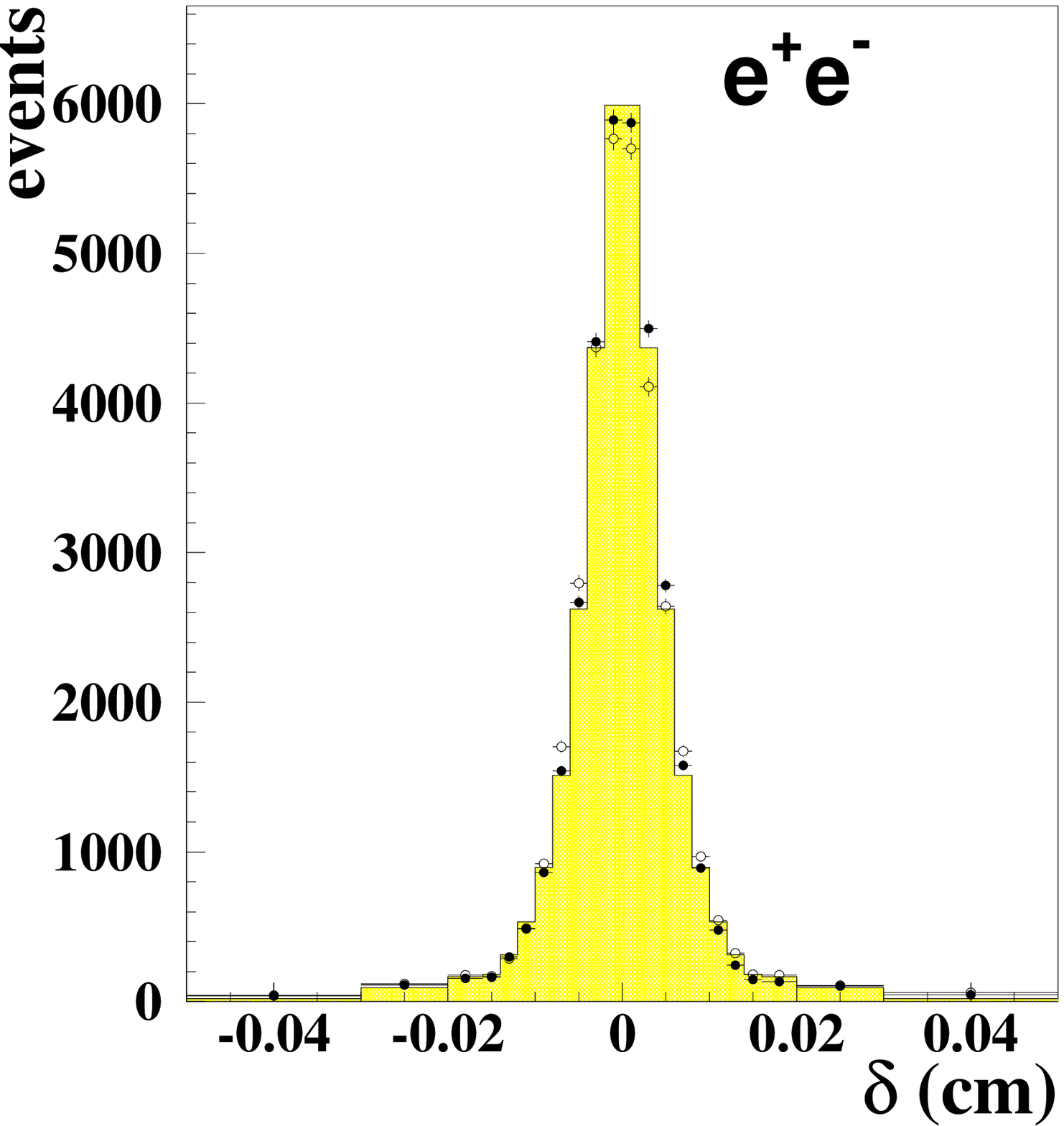,width=0.45\textwidth}
\epsfig{file=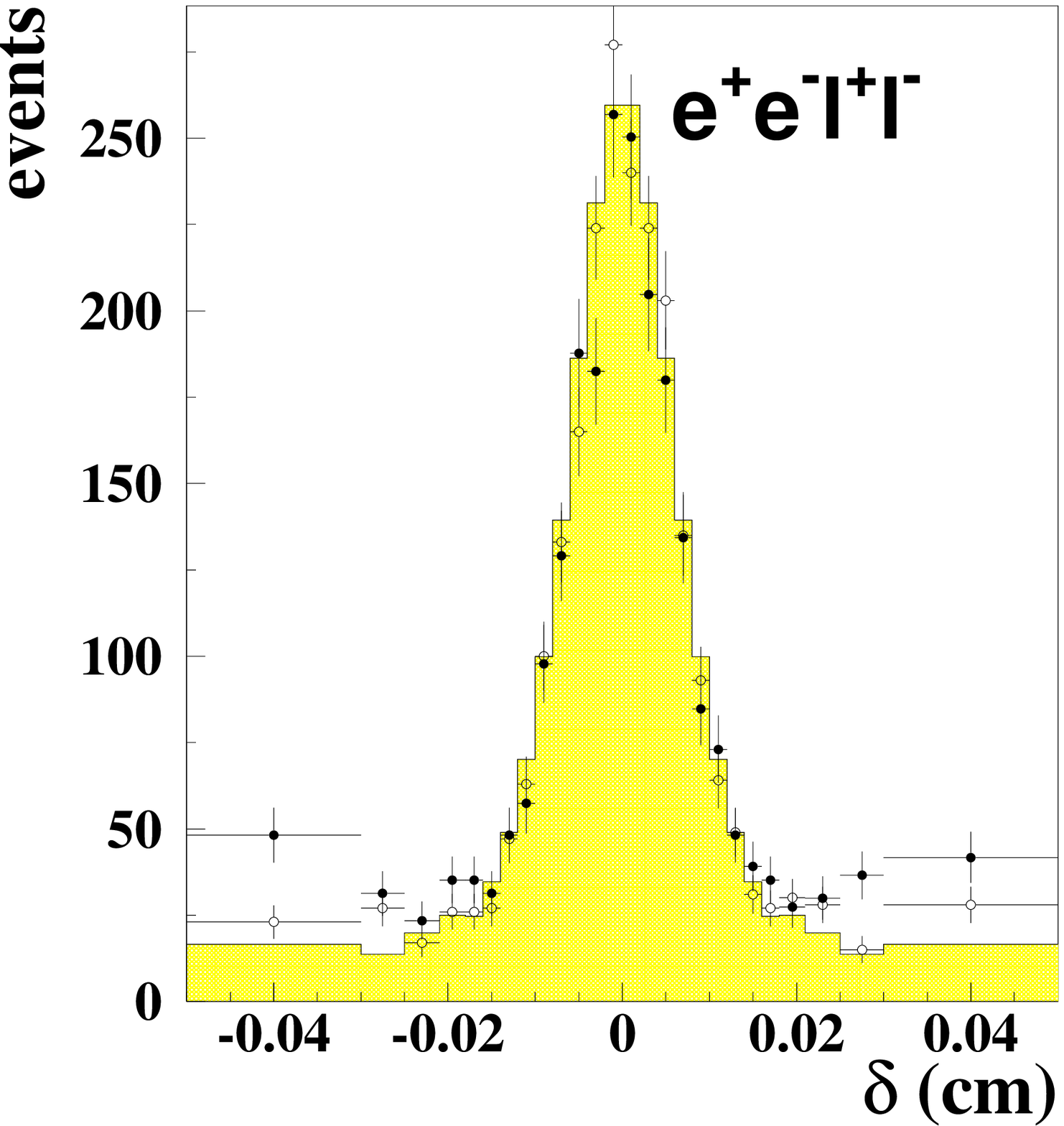,width=0.45\textwidth}
\end{center}
\caption{
Distribution of the miss distance for the calibration samples of dimuon, 
dielectron
and two-photon events from 1994 data ($\circ$) compared with simulation 
($\bullet$) and the
parameterization used (histogram).
}
\label{f:calib94}
\end{figure}

The resolution function was calibrated on the data. This required event 
samples with a topology similar to that of one-prong tau 
decays but
with no lifetime effect. Applying the same track quality cuts and lepton 
identification as for the tau pair selection, but appropriately choosing
the $P_{\mbox{\scriptsize rad}}$ and $E_{\mbox{\scriptsize rad}}$ cuts,
it was possible to select samples with 
e$^+$e$^-$ and $\mu^+\mu^-$ pairs, produced by e$^+$e$^-$ 
annihilation at high energy, and by $\gamma\gamma$ interactions at low energy.
The obtained  purities where 98.2\%\ for the high energy e$^+$e$^-$ pairs, 
99.4\%\ for the high energy $\mu^+\mu^-$ pairs and 94\%\  for the low energy 
lepton pairs.
The approach chosen was to use the data for the estimation of the
resolution at high and low momenta and the simulation to interpolate 
in the intermediate momentum region.
The parameterization used was
\begin{equation}
\begin{array}{lcllcl}
f_1      & = & (1-\alpha) f_{mh} f_{th}     + \alpha f_{ml} f_{tl},     &
\sigma_1 & = & [(1-\alpha)k_{1h} +\alpha k_{1l}]\sigma_\msub{0},    \\
f_2      & = & (1-\alpha) (1-f_{mh}) f_{th} + \alpha (1-f_{ml}) f_{tl}, &
\sigma_2 & = & [(1-\alpha)k_{2h} +\alpha k_{2l}]\sigma_\msub{0},    \\
f_3      & = & (1-\alpha) (1-f_{th})        + \alpha (1-f_{tl}),        &
\sigma_3 & = & [(1-\beta) k_{3h} +\beta  k_{3l}]\sigma_\msub{0},    \\
\end{array}
\end{equation}
where
\begin{equation}
\alpha = \left[ 1 - \frac{p\sin^{3/2}\theta}{E_\msub{beam}}
         \left( 2 - \frac{p\sin^{3/2}\theta}{E_\msub{beam}} \right)
         \right]^a,
\end{equation}
\begin{equation}
\beta  = \left[ 1 - \frac{p\sin^{3/2}\theta}{E_\msub{beam}}
         \left( 2 - \frac{p\sin^{3/2}\theta}{E_\msub{beam}} \right)
         \right]^b.
\end{equation}
and $\sigma_\msub{0}$ is the uncertainty estimation from the b-tagging 
package~\cite{gammabb}.
The parameters $f_{mh}$, $f_{th}$, $k_{1 h}$, $k_{2h}$, $k_{3h}$ and
$f_{ml}$, $f_{tl}$, $k_{1l}$, $k_{2l}$, $k_{3l}$ were determined from the miss
distance distribution at high and low momentum respectively. 
As an example the fitted distributions for the 1994 data sample are 
shown in Figure \ref{f:calib94}. 
The parameters $\alpha$ and $\beta$ range from 1 at low momentum to 0 at high
momentum. The steepness of the change was controlled by the parameters $a$ and
$b$, derived from the simulation.
For hadrons an exponential contribution was added in order
to take into account the effect of elastic hadronic interactions. 

The lifetime was determined by an unbinned maximum likelihood
fit to the observed distribution, where
the probability density function was
given by the convolution of the physics function and the
resolution function described above.
The entire procedure was tested on simulated samples 
from which a bias of $-0.2~\pm~0.9$~fs was measured.

\begin{figure}
\begin{center}
\epsfig{file=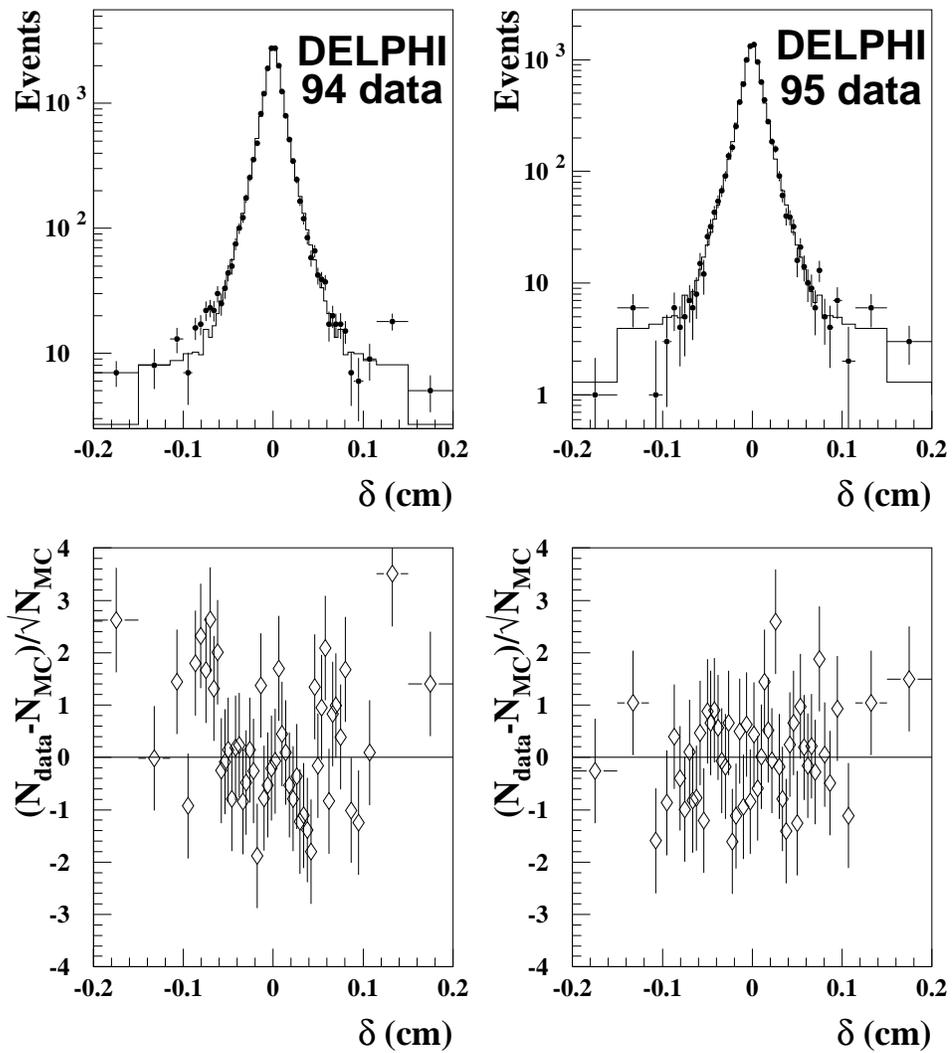,width=\textwidth}
\end{center}
\caption{Distributions of the miss distance for the 1994 and 1995 data 
 samples, points with error bars are data, the histogram the fitted 
 distribution.
 The lower plots show the pulls of the differences between the observed 
 data and the expected events from the simulation.
}
\label{f:mdist9495}
\end{figure}

Fig.~\ref{f:mdist9495} shows the joint distribution of
the miss distance for 1994 and 1995 data, with the best fit 
superimposed.
The measured lifetimes, including all the corrections, are
\begin{eqnarray}
\tau_\tau\mbox{(MD, 94)} &=&
   292.5 \pm 2.8_{stat}\pm 2.3_{sys} \mbox{\ fs}, \nonumber \\
\tau_\tau\mbox{(MD, 95)} &=&
   291.0 \pm 4.0_{stat}\pm 2.3_{sys} \mbox{\ fs}. \nonumber
\end{eqnarray}

The sources of systematic uncertainties are listed in 
Table~\ref{t:MD_sys}.
The event selection criteria were varied inducing a lifetime change
of 1.1 fs in 1994 and 1.0 fs in 1995.
The influence of the physics function and resolution function was checked by
varying by $\pm 1 \sigma$ the parameters of the functions
parameterization, taking into account correlations.
A further contribution to the resolution function,
due to hadronic scattering, was evaluated comparing
the data and simulation for hadronic events.
The residual lepton misidentification, after applying the procedure in 
section~\ref{s:leptid} resulted in a systematic uncertainty of 
$\pm 0.2$~fs.
The effect of background from 
e$^+$e$^-$, $\mu^+\mu^-$ and $\gamma\gamma$ events 
was evaluated using simulated samples that passed all the selection cuts, 
resulting in estimated biases of $-0.6 \pm 0.4$ fs in 1994 
and $-0.8 \pm 0.4$ fs in 1995.
The contribution to the systematic uncertainty due to the alignment 
of the Microvertex Detector, estimated using the same 
procedure as in section~\ref{sec:ipd}, was 0.5~fs.
The dependence on the tau polarization $P_\tau$ was checked by varying 
it in the range
$[-0.11;-0.17]$ resulting in a $\pm 0.1$~fs variation on the lifetime,
while the effect of the transverse polarisation correlation was evaluated as
$\pm 0.4$ fs ~\cite{DELPHI95}. 
The fit was performed over the range $|\delta|<1.5$~mm.
This range was varied between  $|\delta|<1$~mm and  $|\delta|<2$~mm
to study the systematic effect. The maximum difference observed was taken
as a systematic uncertainty.

\begin{table}[h]
\vspace {0.3cm}
\begin{center}
\begin{tabular}{|l|r|r|r|} \hline
                                  & 1994                 &  1995               & Correlation  \\
                                  & [fs]                 &   [fs]              & coefficient  \\  \hline
Fit result                        &  $291.7\pm 2.8$      &   $290.0\pm 4.0$    &    0.0      \\ \hline
Syst. Sources:                    &                      &                     &        \\  
Method bias                       &  $ +0.2\pm 0.9$      &   $ +0.2\pm 0.9$    &    1.0      \\
Event Selection                   &  $\pm1.1$            &   $\pm1.0$          &    0.0       \\
Physics Function                  &  $\pm0.8$            &   $\pm0.8$          &    1.0       \\
Resolution Function               &  $\pm1.3$            &   $\pm1.5$          &    0.9       \\
Particle Misidentification        &  $\pm0.2$            &   $\pm0.2$          &    1.0       \\
Background                        &  $+0.6 \pm 0.4$      &   $+0.8 \pm 0.4$    &    0.7       \\
Alignment                         &  $\pm0.5$            &   $\pm0.5$          &    0.0       \\
Polarisation                      &  $\pm0.4$            &   $\pm0.4$          &    1.0       \\
Fit Range                         &  $\pm0.7$            &   $\pm0.2$          &    0.0       \\ \hline
Result                          & $292.5\pm 2.8\pm 2.3$  & $291.0\pm 4.0 \pm 2.3$ &  0.17     \\ \hline
{\bf Average 94+95}&\multicolumn{3}{c|}{$\mathbf 292.0\pm 2.3\pm 2.1$} \\
\hline
\end{tabular}
\caption[Systematic Uncertainties]{
Summary of results, correction for biases, systematic uncertainties and 
combination of 
the Miss Distance measurements. }
\label{t:MD_sys}
\end{center}
\end{table}

The measurements for the two years were combined,
accounting for the correlations in
the systematic uncertainties shown in Table~\ref{t:MD_sys}, to give the
result
$$ \tau\mbox{(MD, 94-95)} = 292.0 \pm 2.3_{stat} \pm 2.1_{sys}\mbox{\ fs.}$$
                                                                   
\section{Combination of Measurements}
\label{sec:correlations}
 
The lifetime of the tau has been measured with three methods.
The two one-prong measurements were performed on the same data sample and 
were combined taking into account correlated statistical and systematic 
uncertainties.
The statistical correlation was obtained by subdividing the simulated data 
into 89 samples of 5000 events, applying the two analysis methods on each 
sample and computing the correlation coefficient as:
\begin{equation}
\rho = \frac{ \sum_i(\tau_{\msub{IPD},i}-\tau_\tau)
                    (\tau_{\msub{MD},i} -\tau_\tau)
          }{\sqrt{
              \sum_i(\tau_{\msub{IPD},i}-\tau_\tau)^2
              \sum_i(\tau_{\msub{MD},i} -\tau_\tau)^2
          }}, \label{eq:comprho}
\end{equation}
where $\tau_{\msub{IPD},i}$ and $\tau_{\msub{MD},i}$ are the determined 
lifetimes for each sample respectively with the impact parameter difference
and the miss distance methods, and $\tau_\tau$ is the simulated lifetime.
The resulting statistical correlation was  36\%.
Among the systematic
uncertainties only the background estimation and alignment contribution
were correlated. This provides a combined result of   
\begin{equation}
\tau_\tau\mbox{(1-prong, 94+95)} = 291.8 \pm 2.3_{stat}\pm 1.5_{sys} \mbox{\ fs.} \nonumber 
\end{equation}

This measurement was averaged with previously published DELPHI results~\cite{DELPHI95,DELPHI91}:
\begin{eqnarray}
\tau_\tau \mbox{(1-prong, 91)} &=& 298\pm 7_{stat}\pm 4_{sys} \mbox{\ fs}, \nonumber \\
\tau_\tau \mbox{(1-prong, 92--93)} &=& 291.8\pm 3.3_{stat}\pm 2.1_{sys} \mbox{\ fs}, \nonumber
\end{eqnarray}
to provide the result for all LEP-1 DELPHI data for the one-prong
methods:
\begin{equation}
\tau_\tau \mbox{(1-prong, 91-95)} = 292.3\pm 1.8_{stat}\pm 1.2_{sys} \mbox{\ fs}. \nonumber 
\end{equation}

  \label{sec:combi}

The final one-prong estimation was combined with the three-prong measurement
to give the best estimation of the tau lifetime from the DELPHI data. 
Only
the systematic uncertainty attributed to the alignment of the vertex detector
is common between the one-prong and three-prong
measurements, resulting in a  5\% correlation between the two results.
By combining the two results and taking into account this correlation,
a tau lifetime of
\[ \tau_{\tau}  =  290.9 \pm 1.4_{stat} \pm 1.0_{sys} \mbox{\ fs} \]
was obtained.

\section{Summary and Conclusions}
  \label{sec:summary}
The tau lifetime has been measured using the DELPHI LEP-1 data sample.
The result 
\[ \tau_{\tau}  =  290.9 \pm 1.4_{stat} \pm 1.0_{sys} \mbox{\ fs} \]
was obtained.
This result supersedes all previous DELPHI measurements of the tau lifetime.
The measurement is compatible with the values published
by other experiments~\cite{OPAL93} and has a slightly better 
precision.

\par
Tests of  $\tau - \mu $ and $\tau - {\rm{e}} $ universality
can be performed using this result in conjunction with the published DELPHI
values for the $\tau$ leptonic branching fractions \cite{BRpap}:
$$
B(\TEL) = (17.877 \pm 0.109_{stat} \pm 0.110_{sys})\%,
$$
$$
B(\TMU) = (17.325 \pm 0.095_{stat} \pm 0.077_{sys})\%,
$$
together with the world average values of the lepton masses and the
muon lifetime \cite{PDG02}. Using equations (\ref{eqn:life}) and 
(\ref{eqn:life1}), and accounting for small 
radiative corrections
(see \cite{BRpap} for a discussion), yielded
$$
\frac{g_{\tau}} {g_{\mu}} = 1.0015 \pm 0.0053,
$$
$$
\frac{g_{\tau}} {g_{e}} = 0.9997 \pm 0.0046.
$$
The branching fraction measurements contributed to the uncertainty
in these estimates with  $\pm 0.0043 $ 
and $\pm 0.0035 $ respectively.  

\par
\begin{figure}
\begin{center}
\epsfig{file=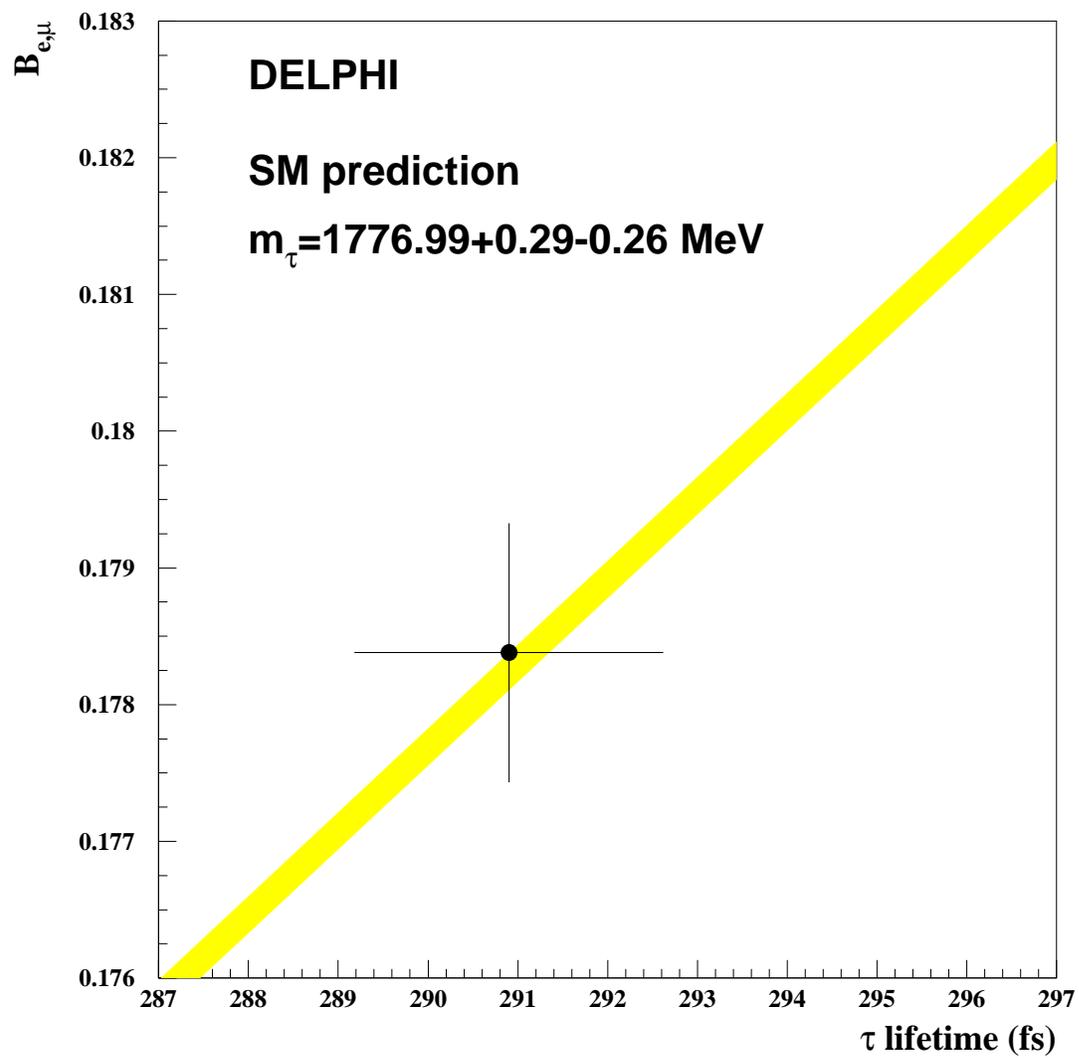,width=\textwidth}
\end{center}
\caption{The measured values of the leptonic branching ratio 
and the lifetime of the tau lepton as measured by DELPHI (point), 
compared to the Standard Model 
relationship (shaded band). 
}
\label{f:univ}
\end{figure}

Under the assumption of e-$\mu$ universality,  $g_{\mu} = g_{e} \equiv 
g_{{\rm e},\mu} $,
it was possible to give a more stringent test of universality of the 
coupling of the $\tau$ and that of the two lighter leptons. 
The two measurements were combined into one leptonic branching
fraction, $B_{ e,\mu}$, 
correcting for the phase space suppression of $B(\TMU) $:
$$ B_{e,\mu} = (17.838 \pm 0.066_{stat} \pm 0.068_{sys})\%,  $$
to compare the  tau  charged current coupling to that of the two
lighter leptons.
The result 
$$ 
\frac{g_{\tau}} {g_{{\rm e},\mu}} = 1.0004 \pm 0.0041 $$
was obtained, in excellent agreement with $\tau$-$(e,\mu)$ universality. 
The relation 
between the leptonic branching ratio and the $\tau$ lifetime is
shown in Fig.~\ref{f:univ}, under the assumption of $e$-$\mu$ universality.
The band reflects the uncertainty on the tau mass \cite{PDG02}.

\subsection*{Acknowledgements}
\vskip 3 mm
 We are greatly indebted to our technical 
collaborators, to the members of the CERN-SL Division for the excellent 
performance of the LEP collider, and to the funding agencies for their
support in building and operating the DELPHI detector.\\
We acknowledge in particular the support of \\
Austrian Federal Ministry of Education, Science and Culture,
GZ 616.364/2-III/2a/98, \\
FNRS--FWO, Flanders Institute to encourage scientific and technological 
research in the industry (IWT), Belgium,  \\
FINEP, CNPq, CAPES, FUJB and FAPERJ, Brazil, \\
Czech Ministry of Industry and Trade, GA CR 202/99/1362,\\
Commission of the European Communities (DG XII), \\
Direction des Sciences de la Mati$\grave{\mbox{\rm e}}$re, CEA, France, \\
Bundesministerium f$\ddot{\mbox{\rm u}}$r Bildung, Wissenschaft, Forschung 
und Technologie, Germany,\\
General Secretariat for Research and Technology, Greece, \\
National Science Foundation (NWO) and Foundation for Research on Matter (FOM),
The Netherlands, \\
Norwegian Research Council,  \\
State Committee for Scientific Research, Poland, SPUB-M/CERN/PO3/DZ296/2000,
SPUB-M/CERN/PO3/DZ297/2000, 2P03B 104 19 and 2P03B 69 23(2002-2004)\\
FCT - Funda\c{c}\~ao para a Ci\^encia e Tecnologia, Portugal, \\
Vedecka grantova agentura MS SR, Slovakia, Nr. 95/5195/134, \\
Ministry of Science and Technology of the Republic of Slovenia, \\
CICYT, Spain, AEN99-0950 and AEN99-0761,  \\
The Swedish Natural Science Research Council,      \\
Particle Physics and Astronomy Research Council, UK, \\
Department of Energy, USA, DE-FG02-01ER41155. \\
EEC RTN contract HPRN-CT-00292-2002. \\



\newpage 


%

\end{document}